\documentclass[prd,aps,twocolumn,a4paper,showkeys,nofootinbib,superscriptaddress,
 reprint,floatfix,
nofootinbib,
 amsmath,amssymb,
 aps
]{revtex4-2}

\usepackage{graphicx}
\usepackage{dcolumn}
\usepackage{bm}
\usepackage[english]{babel}
\usepackage{multirow}
\usepackage{subfiles}
\usepackage{physics}
\usepackage{mathrsfs}
\usepackage[dvipsnames]{xcolor}
\usepackage{booktabs}
\usepackage[normalem]{ulem}

\definecolor{color1}{RGB}{48,92,222} 
\definecolor{color2}{RGB}{0,20,20} 
\usepackage[linktocpage]{hyperref}
\hypersetup{
	hidelinks,
	colorlinks,
	breaklinks=true,
	urlcolor=color2,
	citecolor=color1,
	linkcolor=color1,
	bookmarksopen=false,
	pdftitle={Title},
	pdfauthor={Author},
}

\begin{document}

\preprint{}

\title{Precession resonances in the strong-gravity environment of a Kerr black hole}

\author{Davide Panella}
\email{davide.panella@nbi.ku.dk}
  \affiliation{Dipartimento di Fisica e Geologia, Universit\`a di Perugia, I.N.F.N. Sezione di Perugia, \\ Via Pascoli, I-06123 Perugia, Italy}
 \affiliation{Center of Gravity, Niels Bohr Institute, Copenhagen University,\\ Blegdamsvej 17, DK-2100 Copenhagen \O{}, Denmark}

\author{Francesco Marzocco}
\email{francesco.marzocco@dottorandi.unipg.it}
  \affiliation{Dipartimento di Fisica e Geologia, Universit\`a di Perugia, I.N.F.N. Sezione di Perugia, \\ Via Pascoli, I-06123 Perugia, Italy}

\author{Marta Cocco}
\email{marta.cocco@um.es}
  \affiliation{Dipartimento di Fisica e Geologia, Universit\`a di Perugia, I.N.F.N. Sezione di Perugia, \\ Via Pascoli, I-06123 Perugia, Italy}
  \affiliation{Center of Gravity, Niels Bohr Institute, Copenhagen University,\\ Blegdamsvej 17, DK-2100 Copenhagen \O{}, Denmark}
  \affiliation{Departamento de F\'isica, Universidad de Murcia, Campus de Espinardo, 30100 Murcia, Spain}
\author{Gianluca Grignani}
\email{gianluca.grignani@unipg.it}
\affiliation{Dipartimento di Fisica e Geologia, Universit\`a di Perugia, I.N.F.N. Sezione di Perugia, \\ Via Pascoli, I-06123 Perugia, Italy}
\author{Troels Harmark}
\email{harmark@nbi.ku.dk}
 \affiliation{Center of Gravity, Niels Bohr Institute, Copenhagen University,\\ Blegdamsvej 17, DK-2100 Copenhagen \O{}, Denmark}
  \affiliation{Nordita, KTH Royal Institute of Technology and Stockholm University,\\
  Hannes Alfv\'ens v\"ag 12, SE-106 91 Stockholm, Sweden}

\author{Marta Orselli}
\email{marta.orselli@unipg.it}
 \affiliation{Dipartimento di Fisica e Geologia, Universit\`a di Perugia, I.N.F.N. Sezione di Perugia, \\ Via Pascoli, I-06123 Perugia, Italy}
 \affiliation{Center of Gravity, Niels Bohr Institute, Copenhagen University,\\ Blegdamsvej 17, DK-2100 Copenhagen \O{}, Denmark}

\date{\today}

\begin{abstract}
Precession resonances in b-EMRI systems provide characteristic signatures of the interaction between a compact binary and the supermassive black hole it orbits. In strong gravity, relativistic orbital frequencies allow additional resonances that are absent in the Newtonian regime. In this work, we investigate these resonances for a compact binary orbiting a rotating supermassive black hole described by the Kerr metric. This is an astrophysically relevant scenario, since supermassive black holes are expected to carry angular momentum. We consider both equatorial and inclined outer orbits and include both electric and magnetic tidal interactions. For inclined orbits, the additional polar degree of freedom gives rise to new resonant combinations that are absent in the spherically symmetric case. We derive the corresponding resonance spectrum analytically in the quasi-circular and quasi-equatorial regime, study its dependence on the black hole spin, and test these predictions through numerical integration of the equations of motion. Finally, we identify configurations in which the resonant effects occur while the binary system emits gravitational waves in the LISA frequency band.
\end{abstract}

\maketitle

\tableofcontents

\section{\label{Sec:Introduction}Introduction}
In recent years, binary extreme mass-ratio inspirals (b-EMRIs) have been proposed as realistic astrophysical systems~\cite{Chen:2018axp,Generozov:2018niv}. A b-EMRI consists of a bound pair of compact objects, which may be two black holes (BHs), two neutron stars, or a black hole and a neutron star, orbiting a supermassive black hole (SMBH).
Several formation channels for b-EMRIs have been proposed in dense galactic environments. In particular, galactic nuclei hosting SMBHs can capture pre-existing compact binaries through their gravitational potential~\cite{Hopman:2006qr,Chen:2018axp,Generozov:2018niv}. Compact binaries can also form through dynamical interactions~\cite{OLeary:2008myb,Samsing:2017rat}. In active galactic nuclei (AGNs), migration traps provide an additional mechanism for the formation of b-EMRIs~\cite{Bellovary:2015ifg,Secunda:2020mhd,Peng:2021vzr,Tagawa:2019osr}. The GW190521 event~\cite{LIGOScientific:2020iuh}, corresponding to the merger of two stellar-mass black holes, has for instance been interpreted as a possible merger occurring within an AGN migration trap~\cite{Toubiana:2020drf,Graham:2020gwr,Sberna:2022qbn}. Migration processes may also account for the presence of a stellar-mass binary in the vicinity of Sgr A$^*$~\cite{Peissker:2024ade}. These different mechanisms support the astrophysical existence of b-EMRIs in dense galactic environments.

From an observational perspective, current interferometers have enabled the detection of gravitational-wave (GW) signals from compact-object binary mergers~\cite{LIGOScientific:2016dsl,LIGOScientific:2018mvr,LIGOScientific:2020ibl}. Future-generation detectors are expected to extend the accessible frequency band, in particular towards lower frequencies (mHz)~\cite{Amaro-Seoane:2012aqc,amaroseoane2017laserinterferometerspaceantenna,LISA:2022yao,Punturo:2010zz,Barack:2009ux,Berry:2019wgg}, thereby allowing observations over timescales significantly longer than those currently available.
In this context, b-EMRIs could represent particularly informative sources, given the large number of orbital cycles they can complete prior to merger~\cite{amaroseoane2017laserinterferometerspaceantenna}. Their observation could provide information on their formation channels, dynamical evolution, environmental effects, and the combined imprint of these contributions on the emitted waveform~\cite{Camilloni:2023xvf,Hendriks:2024gpp,Jiang:2024mdl,Santos:2025ass,Santos:2026lzq,Yin:2024nyz,Katagiri:2026gkz,Cardoso:2026ugm,Meng:2024kug, Seymour:2026txk}. In b-EMRIs, relativistic precession, Kerr orbital dynamics, and tidal interactions can leave cumulative imprints on the GW signal emitted by the inner binary.

In a b-EMRI, the two compact objects form an inner binary, while their center of mass follows an outer orbit around the SMBH. In such systems, tidal forces~\cite{Antonini:2012ad,Samsing:2017rat,Samsing:2017xmd} exerted by the SMBH on the inner binary play a crucial role in shaping its orbital evolution and long-term dynamics. Taking the outer orbital period as the reference timescale, both secular and non-secular effects associated with tidal interactions can arise.

In this work, we focus on the non-secular phenomenon of \textit{precession resonances}, which occur when the precession frequency of the inner binary becomes commensurate with a linear combination of the characteristic frequencies associated with the outer motion.

It was first shown in Ref.~\cite{Kuntz:2021hhm} for the case of a Newtonian system that, when non-secular effects are taken into account, the precession frequency of the inner orbit can become commensurate with the Keplerian orbital frequency of the outer motion, giving rise to precession resonances.
As the binary inspirals due to gravitational radiation-reaction (RR), its precession frequency increases, allowing the system to pass through a sequence of resonances.
When a resonance condition is met, an otherwise oscillatory contribution accumulates coherently and drives changes in the eccentricity of the inner binary and in other orbital elements. The resulting eccentricity excitation modifies the subsequent RR-driven evolution and, consequently, the emitted GW signal.

The study of precession resonances was later extended to the relativistic regime in Ref.~\cite{Cocco:2025adu}, where the inner binary was considered to be embedded in the tidal field of a Schwarzschild black hole. Several important differences arise with respect to the Newtonian description. First, the tidal interaction is modeled perturbatively in terms of relativistic quadrupolar electric tidal moments, which generalize their Newtonian counterparts to the strong-field regime. Second, the degeneracy of the Keplerian orbital frequency is broken, giving rise to two distinct relativistic frequencies and thereby enriching the spectrum of possible resonances. The quadrupolar-order analysis was completed in Ref.~\cite{Cocco:2025udb} by including the magnetic tidal interaction. Magnetic tidal moments have no Newtonian counterpart and enter the dynamics at 0.5 post-Newtonian (PN) order. Consequently, the resonant effects associated with the magnetic tidal coupling provide intrinsically relativistic contributions to the dynamics and to the emitted GW signal.

The analysis performed in Refs.~\cite{Cocco:2025adu, Cocco:2025udb} revealed that precession resonances could be used as a probe of strong-gravity effects by analyzing the imprint that resonant peaks in the eccentricity evolution of the inner orbit can potentially leave on the emitted GW signal. This signature may be accessible to future GW detectors, such as LISA.

Since astrophysical SMBHs are expected to rotate, we now extend the analysis of precession resonances to the Kerr geometry.
Compared with a spherically symmetric background, the axial symmetry of the Kerr geometry introduces additional orbital degrees of freedom, including non-trivial polar motion for non-equatorial outer orbits. This richer dynamics also allows us to investigate how the spin of the SMBH affects the resonances.

One of our main results is the derivation of the following resonance condition
\begin{equation}
\label{introresonancecondition}
q\dot{\gamma}=k\Omega_{\hat{r}}+l\Omega_{\hat{z}}+m\Omega_{\hat{\Psi}}~,
\end{equation}
where $\dot{\gamma}$ is the precession frequency of the inner binary, $\Omega_{\hat{r}}$ and $\Omega_{\hat{z}}$ are the radial and polar frequencies of the outer orbit, respectively, and $\Omega_{\hat{\Psi}}$ is the frequency associated with Marck's angle $\hat{\Psi}$, which describes the rotation of the parallel-transported tetrad along the outer orbit~\cite{03e6e95f-6105-3f53-9aa8-4864f05e2d86}. Finally, $q$, $k$, $l$, and $m$ are integers. The presence of the polar frequency further enriches the spectrum of possible resonances, introducing new resonant combinations that are absent for equatorial motion. For astrophysically relevant configurations, some of these resonant effects can occur while the inner binary emits gravitational waves in the LISA frequency band.

In this work, we identify the precession resonances arising from the condition~\eqref{introresonancecondition} when both the leading electric (0PN) and magnetic (0.5PN) tidal interactions are included. In particular, we include in our analysis both equatorial and off-equatorial orbits in the Kerr spacetime. We investigate the new resonant structures associated with the polar degree of freedom and how the spin of the SMBH affects these resonances.

The paper is organized as follows. In Sec.~\ref{Sec:Assumptions}, we outline our assumptions on the masses, separations, and characteristic timescales of the system, which ensure both physical consistency and analytical control. In Sec.~\ref{Sec:Hamiltonian}, we review the Hamiltonian describing the dynamics of the system and derive the equations of motion. In Sec.~\ref{Sec:Fourier}, we analytically investigate the dynamics of the inner orbit using a perturbative treatment of quasi-circular and quasi-equatorial outer orbits. This allows us to perform a Fourier expansion of the interaction Hamiltonian, identify the resonances arising at the chosen perturbative order, and derive the corresponding Fourier coefficients, including their dependence on the black hole spin. In Sec.~\ref{Sec:Numerical}, we test the analytical predictions numerically by integrating the equations of motion for different orbital configurations and identifying the predicted resonances in the eccentricity evolution of the inner orbit. We first consider equatorial prograde and retrograde geodesics, thereby isolating the effect of the black hole spin on the dynamics. We then study off-equatorial orbits, highlighting the emergence of new resonances associated with the polar degree of freedom, and discuss their potential observational implications. Finally, in Sec.~\ref{Sec:Conclusions}, we summarize our findings and discuss possible directions for future work.

To guide the reader through the text, we adopt the convention that, whenever the same symbol is used for both the inner and outer motions, hatted quantities refer to the outer orbit, whereas unhatted quantities refer to the inner binary. Unless otherwise specified for the mode indices $(k,l,m)$, Latin indices run from 1 to 3, while Greek indices run from 0 to 3. We use the mostly-plus metric signature, $(-,+,+,+)$. Throughout the paper, we do not adopt geometrized units.


\section{Setup and assumptions} \label{Sec:Assumptions}

In this section, we introduce the b-EMRI system and specify the physical regime of interest for this work.

We consider a binary system (hereafter referred to as the \textit{inner binary}) consisting of two compact objects with masses $m_1$ and $m_2$ in a bound orbit.
The inner binary has total mass $M=m_1+m_2$, reduced mass $\mu=m_1m_2/M$, and orbital separation $r$.
The two compact objects may be stellar-mass black holes, neutron stars, or one of each.
The center of mass (CoM) of the inner binary follows a bound orbit around a supermassive black hole (SMBH) of mass $M_*$, with radial coordinate $\hat{r}$, thereby forming the \textit{outer binary}.

We first review the assumptions concerning the masses and the orbital separations, and then turn to the relevant timescales.
As we shall see, these two aspects are closely related.


\subsection{\label{Sec: Masses and separations} Hierarchical configuration}

The interaction between the SMBH and the inner binary is treated perturbatively within the \textit{small-tide approximation}. This approximation assumes that the characteristic size $r$ of the inner binary is much smaller than the curvature radius $\mathscr{R}$ associated with the spacetime generated by the SMBH, \textit{i.e.} $r \ll \mathscr{R}$~\cite{Camilloni:2023xvf}.
We retain only the leading contribution, of order $\mathscr{O}\left((r/\mathscr{R})^2\right)$, corresponding to the quadrupolar approximation for the tidal interaction between the SMBH and the inner binary.
The small-tide approximation can be realized in two distinct regimes~\cite{Poisson:2009qj}: (a) the \textit{weak-field approximation} (small velocities), in which, at sufficiently large distances from the SMBH, the tidal interaction can be described in the Newtonian regime; and (b) the \textit{small-hole approximation} (mass hierarchy), which relies on the mass hierarchy $m_1,m_2 \ll M_*$.
We focus on the latter regime, in which the CoM of the inner binary can probe the strong-field region of the Kerr SMBH, allowing us to study precession resonances in strong gravity.

In principle, the outer orbital dynamics receives gravitational self-force corrections at order $M/M_*$, including both conservative and dissipative effects.
In our treatment, we neglect these corrections and approximate the CoM motion by a Kerr geodesic. In particular, the hierarchy of timescales discussed in Sec.~\ref{Sec: Timescales} justifies neglecting the dissipative evolution of the outer orbit over the timescale of interest.

It is also important to emphasize that we focus on bound outer orbits around the Kerr SMBH.
Therefore, although the small-tide approximation allows the inner binary to approach the event horizon of the SMBH, we must ensure that the orbital parameters of its CoM lie on the stable side of the separatrix, which defines the innermost stable bound orbit (ISO) for a given eccentricity and inclination.\footnote{Closed-form analytic expressions of the separatrix exist in the Schwarzschild limit and for equatorial trajectories in Kerr spacetime~\cite{Cutler:1994pb,Stein:2019buj}, while for generic orbits it is typically determined numerically. A review of the topic can be found, \textit{e.g.}, in Refs.~\cite{Glampedakis:2002ya,Stein:2019buj}.}

We must also ensure that the inner binary remains stable against tidal disruption by the SMBH.
For a given orbital configuration, tidal forces may become strong enough to overcome the binding energy of the inner binary and disrupt it.
Since we aim to follow the evolution of the inner binary from inspiral to merger, we exclude this possibility by imposing the condition
\begin{equation}
\label{stabilityagainstTDE}
    a\lesssim\frac{1}{4}\hat{a}(1-\hat{e})
    \left(\frac{M}{3M_*}\right)^{1/3}\,,
\end{equation}
where $a$ is the inner semi-major axis, while $\hat{a}$ and $\hat{e}$ are the outer semi-major axis and eccentricity, respectively.
This corresponds to the classical Hill criterion for the stability of a system against tidal forces~\cite{Antonini:2012ad, Camilloni:2023xvf, Cocco:2025adu}, extended to relativistic regimes by tightening the classical upper bound by a factor of $1/4$~\cite{Suzuki:2020vfw, Santos:2025ass}.\footnote{The 1PN analysis of Ref.~\cite{Suzuki:2020vfw} does not include spin-orbit effects. Consequently, the approximate criterion adopted here has no explicit dependence on the spin of the SMBH.}

Having outlined the assumptions for the outer orbit, we now turn to the description of the inner binary.
While the outer motion follows a geodesic in the Kerr spacetime generated by the central SMBH, we describe the internal motion of the inner binary using Newtonian dynamics supplemented by post-Newtonian (PN) corrections.
The PN expansion is valid provided that the separation $r$ between the two compact objects is much larger than their Schwarzschild radii, $r \gg 2Gm_i/c^2$, $i=1,2$.
Within this framework, we can also account for relativistic periastron precession and GW radiation reaction through higher-order PN corrections.


\subsection{\label{Sec: Timescales}Timescales}

Our model relies on a hierarchy of characteristic timescales, which is discussed in detail in Ref.~\cite{Cocco:2025udb}; here we only report a summary.
Several studies of the secular dynamics of hierarchical triple systems consider an average over both the inner and outer orbital motions~\cite{Randall:2018qna, Camilloni:2023xvf}, retaining only effects that occur on timescales longer than both orbital periods, such as the von Zeipel--Lidov--Kozai (ZLK) mechanism~\cite{1910AN....183..345V, Lidov:1962wjn, Kozai:1962zz, Hu:2024hzu,Chandramouli:2021kts, Randall:2018qna,Camilloni:2023xvf,Antonini:2012ad,Naoz:2016cjb,Maeda:2023uyx,Maeda:2023tao,Maeda:2025row,Maeda:2026ozk}.
Precession resonances, instead, arise when the 1PN periastron-precession frequency of the inner binary becomes comparable with the fundamental frequencies of the outer orbit. They are therefore non-secular effects with respect to the outer motion~\cite{Kuntz:2021hhm}. To describe these effects we thus do not average over the outer orbit, and require the 1PN precession timescale $T_{\text{1PN}}\sim 1/\dot\gamma_{\text{1PN}}$ to be comparable with the outer orbital period $T_{\text{out,orb}}$ and much shorter than the ZLK timescale $T_{\text{ZLK}}$, so that standard ZLK oscillations are suppressed even for large initial inclinations~\cite{Naoz:2016cjb, Camilloni:2023xvf, Cocco:2025adu, Cocco:2025udb}.

On the other hand, we average over the inner orbit. This requires the inner orbital period $T_{\text{inn,orb}}$ to be the shortest timescale of the problem, \textit{i.e.}, it is shorter than $T_{\text{1PN}}$, since the 1PN precession is slow compared to the Newtonian orbital motion.
The fact that $T_{\text{inn,orb}}$ is shorter than $T_{\text{out,orb}}$ is due to the hierarchical configuration. Moreover, $T_{\text{inn,orb}}$ is shorter than the timescale $T_{\text{tidal}}$ over which the tidal field varies along the outer orbit. One has $T_{\text{tidal}}\sim T_{\text{out,orb}}$ for quasi-circular orbits, while for eccentric orbits it is set by the duration of the periastron passages~\cite{Cocco:2025udb}.

Finally, the inner binary loses energy and angular momentum through GW emission, which we include by adding the leading 2.5PN dissipative corrections~\cite{Peters:1963ux, Peters:1964zz}. For an initial semi-major axis $a_0$ and initial eccentricity $e_0\ll1$, the corresponding approximate radiation-reaction (RR) timescale $T_{\text{inn,RR}}$ can be found in Ref.~\cite{2007gwte.book.....M}.
On the other hand, we neglect GW emission from the outer orbit, which requires $T_{\text{inn,RR}}$ to be much shorter than the outer RR timescale, $T_{\text{out,RR}}\sim (M_*/M)\,T_{\text{out,orb}}$, where $T_{\text{out,orb}}$ is defined through the azimuthal frequency of the outer orbit~\cite{Hinderer:2008dm,Cocco:2025udb}. In this way, the CoM of the inner binary follows a geodesic of the Kerr spacetime up to merger~\cite{03e6e95f-6105-3f53-9aa8-4864f05e2d86, Camilloni:2023xvf, Kuntz:2021ohi, Kuntz:2022onu}.

Overall, the hierarchy of timescales reads
\begin{equation}
\label{timescales}
    \begin{aligned}
        &T_{\text{inn,orb}}\ll T_{\text{out,orb}}\sim T_{\text{1PN}}\ll T_{\text{ZLK}}\,,\\
        &T_{\text{inn,orb}}\ll T_{\text{tidal}}\,,\\
        &T_{\text{inn,RR}} \ll T_{\rm out,RR}\,.
    \end{aligned}
\end{equation}
The masses and orbital parameters used in Sec.~\ref{Sec:Numerical} are chosen accordingly.


\section{\label{Sec:Hamiltonian}Hamiltonian and equations of motion}

In this section, we briefly review the Hamiltonian describing the dynamics of the binary system in the tidal field of a Kerr black hole, along with the associated equations of motion. We can decompose the Hamiltonian as
\begin{equation}
\label{decompositionHamiltonian}
    H=H_{\text{inn}}+H_{\text{tidal}}+H_{\text{pp}}\,,
\end{equation}
and analyze each contribution separately.
The dynamics of the inner binary is described consistently up to 0.5PN order, as required to include the leading quadrupolar tidal couplings (electric at 0PN and magnetic at 0.5PN). On top of this, we add the leading PN corrections responsible for periastron precession (1PN) and radiation reaction (2.5PN).


\subsection{\label{Sec:Inner Hamiltonian}Inner Hamiltonian}

We have already established that the hierarchy of the timescales under consideration allows for averaging over the inner orbital dynamics. We therefore begin by briefly reviewing the averaging procedure, introducing the orbital parameters required to describe the dynamics and evolution of the inner binary.
The unperturbed Newtonian two-body problem can be conveniently analyzed in the CoM frame~\cite{2002clme.book.....G}, where the motion reduces to that of a fictitious particle with mass $\mu$ moving under the influence of a central potential and described by the relative coordinate $\bm{x}$. In the bound case, the general solution corresponds to a Keplerian ellipse. The radial distance $r = |\bm{x}|$ can be parametrized in terms of the \textit{true anomaly} $\psi$, defined as the angle between the direction of periastron and the instantaneous position of the particle, as follows
\begin{equation}
\label{innerorbitparametrisation}
r = \frac{a(1 - e^2)}{1 + e \cos \psi},
\end{equation}
where $a$ and $e$ denote the \textit{semi-major axis} and \textit{eccentricity} of the orbit, respectively. 

Our setup, however, corresponds to a perturbed two-body problem, as the binary system is subject to the tidal field of an external SMBH. In this case, different constraints can be imposed on the dependence of the solution $\bm{x}$ on the orbital parameters, while still yielding the same physical trajectory $\bm{x}(t)$. In other words, the equations of motion governing the time evolution of the orbital elements exhibit a gauge symmetry. This property can be exploited by adopting the \textit{generalized Lagrange gauge}, which allows one to use the canonical momentum of the unperturbed two-body system $\bm{p}$, conjugate to $\bm{x}$, even in the presence of perturbations, \textit{i.e.}, without explicitly incorporating the tidal interaction with the SMBH into $\bm{p}$~\cite{Efroimsky:2003pg, Efroimsky:2003mz}.

The orientation of the inner binary is characterized by the Euler angles $(\gamma, I, \vartheta)$, defined with respect to the local reference frame comoving with its center of mass. Here, $\gamma$ is the \textit{argument of periapsis}, $I$ the \textit{orbital inclination}
relative to the reference plane, and $\vartheta$ the \textit{longitude of the ascending node}.

At leading Newtonian order, the inner binary follows a periodic Keplerian orbit, making the action--angle formalism a natural choice for describing its dynamics.
A standard choice in celestial mechanics is given by the \textit{Delaunay variables}
\cite{delaunay1860théorie,2002clme.book.....G,2002mcma.book.....M}, for which the angle variables are $(\beta,\gamma,\vartheta)$, with $\beta$ denoting the \textit{mean anomaly}.
The corresponding conjugate action variables
$(J_\beta,J_\gamma,J_\vartheta)$ are given by
\begin{equation}
\label{Delunayconjmomenta}
    J_\beta = \mu \sqrt{G M a}\,, \quad
    J_\gamma = J_\beta \sqrt{1-e^2}\,, \quad
    J_\vartheta = J_\gamma \cos I\,.
\end{equation}
To perform the orbital averaging of a given quantity, we integrate over a full cycle of the mean anomaly $\beta$, since it is the periodic angle that increases uniformly in time during the elliptic motion. For analytical convenience, this average is often computed by performing the integral over the true anomaly $\psi$. To this end, one exploits some well-known relations between $\beta$, $\psi$, and the \textit{eccentric anomaly} $\zeta$, namely
\begin{equation}
\label{kepleriananomalies}
    \cos \psi = \frac{\cos \zeta - e}{1 - e \cos \zeta}\,, \quad \beta = \zeta - e \sin \zeta\,.
\end{equation}
The average of an arbitrary physical quantity $\mathcal{A}$ is then given by
\begin{equation}
\label{averagingprocedure}
    \langle \mathcal{A} \rangle = \frac{1}{2\pi} \int_{0}^{2\pi} \mathcal{A} d\beta = \frac{1}{2\pi} \int_{0}^{2\pi} \mathcal{A} \frac{(1 - e^2)^{3/2}}{(1 + e \cos \psi)^2} d\psi\,.
\end{equation}
It is worth noting that this averaging scheme remains valid as long as the dynamics of the inner binary is accounted for up to 0.5PN order; however, it must be appropriately modified when considering higher-order perturbative corrections~\cite{Kuntz:2021ohi, Kuntz:2022onu}.

In the purely Newtonian case, $H_{\text{inn}}$ corresponds to the binding energy of the binary and, when expressed in Delaunay variables, takes the form~\cite{Randall:2018qna, Camilloni:2023xvf}
\begin{equation}
\label{innerenergy0PN}
    H_{\text{inn,0PN}} = -\frac{G M \mu}{2a}.
\end{equation}
By construction, this term is independent of the true anomaly $\psi$ and therefore does not require any averaging procedure. However, in our analysis, it is crucial to also account for the periastron precession of the inner binary’s orbit. This effect is added as a 1PN correction to the Newtonian inner Hamiltonian and, as obtained from the averaging procedure described above, reads\footnote{The 1PN averaged Hamiltonian for an isolated binary also contains a term depending only on the semi-major axis $a$. This term only contributes to the evolution of the mean anomaly $\beta$. However, since the averaged Hamiltonian does not depend on $\beta$, this equation decouples from the others and does not affect the evolution of the remaining orbital elements.}~\cite{Randall:2018qna, Kuntz:2021hhm}
\begin{equation}
\label{innerenergy1PN}
    \langle H_{\text{inn,1PN}}\rangle = -\mu\frac{3 G^2M^2}{c^2a^2\sqrt{1-e^2}}\,.
\end{equation}
All together, the averaged expression for the inner Hamiltonian is given by $\langle H_{\text{inn}}\rangle =H_{\text{inn,0PN}} +\langle H_{\text{inn,1PN}}\rangle $.


\subsection{\label{sec:Tidal Hamiltonian}Tidal Hamiltonian}

The inner binary moves in the tidal field generated by a rotating SMBH. The Kerr metric for a black hole of mass $M_*$ and angular momentum $J_*$, in modified Boyer-Lindquist coordinates $\hat{x}^\mu = (\hat{t}, \hat{r}, \hat{z}, \hat{\phi})$ takes the form
\begin{equation}
\label{Kerrmetric}
\begin{aligned}
    d\hat{s}^2 = -\left(1 - \frac{2GM_*\hat{r}}{c^2\Sigma}\right)c^2d\hat{t}^2 - \frac{4GM_*\hat{r}}{c^2\Sigma} \mathrm{a} (1-\hat{z}^2) \ cd\hat{t} \, d\hat{\phi} \\
    + \frac{\mathscr{A}}{\Sigma} (1-\hat{z}^2) \ d\hat{\phi}^2 + \frac{\Sigma}{\Delta}d\hat{r}^2 + \frac{\Sigma}{1-\hat{z}^2} d\hat{z}^2 \ ,
\end{aligned}
\end{equation}
where $\mathrm{a} = J_*/(cM_*)$ is the specific angular momentum and
\begin{equation}
\label{Kerrmetricdefinitions}
\begin{gathered}
    \Sigma = \hat{r}^2 + \mathrm{a}^2 \hat{z}^2, \quad \Delta = \hat{r}^2 - 2\frac{GM_*}{c^2} \hat{r}+ \mathrm{a}^2, \\
    \mathscr{A} = (\hat{r}^2 + \mathrm{a}^2)^2 - \mathrm{a}^2 \Delta (1-\hat{z}^2) \ .
\end{gathered}
\end{equation}
The coordinate $\hat{z}$ is related to the usual polar Boyer-Lindquist coordinate $\hat{\theta}$ through $\hat{z}=\cos{\hat{\theta}}$.

The coupling of the inner binary to the tidal field of the Kerr background (\textit{i.e.}, the interaction between the curvature of the background spacetime and the internal, finite-size structure of the inner binary) is encoded in the so-called \textit{tidal moments}, which are components of the \textit{tidal tensor}. 
This tensor can be computed at different perturbative orders, each associated with powers of the ratio between the size $r$ of the object embedded in the external gravitational field and the characteristic length scale $\mathcal{R}$ over which it varies. 
In our case, the small-tide approximation ensures that $r \ll \mathscr{R}$.
Therefore, under this assumption, we retain the quadrupole approximation as the leading-order contribution to the tidal moments. 
At this order, one can construct two symmetric trace-free (STF) tensors, namely the \textit{electric} and \textit{magnetic tidal tensors}, with components $\mathscr{E}_{ij}$ and $\mathscr{B}_{ij}$ (where $i,j=1,2,3$ are spatial indices).
To construct these components one evaluates the \textit{Weyl tensor} $C_{\mu\nu\rho\sigma}$ along Kerr geodesics~\cite{Poisson:2009qj, Camilloni:2023rra}.
This can be achieved by introducing an orthonormal tetrad $\lambda_A^\mu$ (with $A = 0, 1, 2, 3$), which defines a local reference frame at each point along the geodesic. We choose the tetrad, called \textit{Marck's tetrad}~\cite{03e6e95f-6105-3f53-9aa8-4864f05e2d86}, that has the property of being parallel-transported along the geodesic motion~\cite{Camilloni:2023rra}. This feature is realized by introducing a rotation angle, the \textit{Marck's angle} $\hat{\Psi}$, and by enforcing that it satisfies the following evolution equation with respect to the proper time $\hat{\tau}$
\begin{equation}
\label{Marckevolution}
    \frac{d\hat{\Psi}}{d\hat{\tau}} = c\frac{\sqrt{\hat K}}{\Sigma} \!\left( \!\frac{\hat{E}(\hat{r}^2 + \mathrm{a}^2) - \mathrm{a}\hat{L}_{\hat{z}}}{\hat{r}^2 + \hat K} + \mathrm{a} \frac{\hat{L}_{\hat{z}} - \mathrm{a}\hat{E} (1-\hat{z}^2)}{\hat K - \mathrm{a}^2 \hat{z}^2}\! \right) \, ,
\end{equation}
where $\hat{E}$, $\hat{L}_{\hat{z}}$, and $\hat K$ are the constants of motion for timelike geodesics in Kerr spacetime, namely the energy in units of $Mc^2$ (dimensionless), the $\hat z$-component of the angular momentum in units of $Mc$ (with dimensions of length), 
and the \textit{Carter constant} (with the same dimensions as $\hat{L}_{\hat{z}}^2$). Analytic solutions for this first-order differential equation are provided in Refs.~\cite{03e6e95f-6105-3f53-9aa8-4864f05e2d86, vandeMeent:2019cam}, while explicit and practically useful expressions for the conserved quantities can be found in Refs.~\cite{Schmidt:2002qk, vandeMeent:2019cam}.

In order to explicitly compute the electric and magnetic quadrupole moments, we first need the components of the following rank-2 and rank-3 tidal tensors in the basis of the tetrad $\lambda_A^\mu$
\begin{equation}
\label{Projectionontetrad}
\begin{aligned}
    C_{ij}=C_{\mu\nu\rho\sigma}\lambda_0^\mu\lambda_i^\nu \lambda_0^\rho\lambda_j^\sigma\,,\\
    C_{ijk}=C_{\mu\nu\rho\sigma}\lambda_i^\mu\lambda_j^\nu \lambda_k^\rho\lambda_0^\sigma\,,
\end{aligned}
\end{equation}
with $i,j,k=1,2,3$. The electric and magnetic quadrupole moments are then defined as~\cite{Poisson:2009qj}
\begin{equation}
\label{electricmagentictidaldefinition}
    \mathscr{E}_{ij} = C_{ij} \ , \quad \mathscr{B}_{ij} = \frac{1}{2} \epsilon_{ikl} C^{kl}{}_j \ ,
\end{equation}
with $\epsilon_{ijk}$ denoting the three-dimensional Levi-Civita symbol, with $\epsilon_{123}=+1$.\footnote{Notice that, being STF tensors, both the electric $\mathscr{E}_{ij}$ and magnetic $\mathscr{B}_{ij}$ tidal tensors possess five independent components each; together, they account for the ten independent components of the Weyl tensor.} The indices $(i,j,k,\ldots)$ are raised and lowered using the Kronecker delta $\delta_{ij}$. The explicit expressions for the components of the tidal tensors $\mathscr{E}_{ij}$ and $\mathscr{B}_{ij}$ are given in Appendix~\ref{Appendix A: Tidal moments}. 
It is worth noting that, while electric tidal moments appear already at 0PN order and admit a Newtonian counterpart, magnetic tidal moments are purely relativistic contributions entering the dynamics at 0.5PN order.
We point out that our work represents the first application in the literature of the full (off-equatorial) magnetic tidal moments in Kerr spacetime, originally derived in Ref.~\cite{Camilloni:2023rra}.

Electric and magnetic tidal moments encode two distinct types of tidal interactions. 
The corresponding tidal Hamiltonian is~\cite{Cocco:2025udb}
\begin{equation}
\label{Electricandmagnetictidalhamiltonian}
\begin{aligned}
    H_{\mathscr{E}}&=\frac{1}{2}c^2\mathscr{Q}^{ij}\mathscr{E}_{ij}\, ,\\
    H_{\mathscr{B}}&=\frac{2}{3}c\mathscr{T}^{ij}\mathscr{B}_{ij}\,,
\end{aligned}
\end{equation}
where $\mathscr{Q}^{ij}$ and $\mathscr{T}^{ij}$ are rank-2 tensors encoding the internal dynamics of the inner binary system.\footnote{Notice that it is equivalent to use the full or the traceless version of $\mathscr{Q}^{ij}$ and $\mathscr{T}^{ij}$, since they are contracted with STF tidal tensors.} We average the interaction Hamiltonians~\eqref{Electricandmagnetictidalhamiltonian} to remove effects occurring on timescales comparable with the inner orbital period. It is important to emphasize that the averaging procedure is applied only to $\mathscr{Q}^{ij}$ and $\mathscr{T}^{ij}$, since $\mathscr{E}_{ij}$ and $\mathscr{B}_{ij}$ do not depend on the true anomaly $\psi$. The result of this procedure is therefore the same as in Ref.~\cite{Cocco:2025udb}
\begin{align}
\label{Averagedelectricandmagnetictidalhamiltonian1}
    \langle \mathscr{Q}^{ij}\rangle&=\frac{\mu a^2}{2}\left[ (1+4e^2)u^iu^j+(1-e^2)v^iv^j\right]\,,
    \\
\label{Averagedelectricandmagnetictidalhamiltonian2}
    \langle\mathscr{T}^{ij}\rangle&=-\frac{3}{2} \frac{m_1 - m_2}{M} \mu e a \sqrt{G M a (1 - e^2)}\,u^i (\hat{\bm{u}}\cross \hat{\bm{v}})^j\,,
\end{align}
with
\begin{equation}
\label{Unitvectors}
\hat{\bm{u}} = R_z(\vartheta) R_x(I) R_z(\gamma) \hat{\bm{x}} \ , \quad \hat{\bm{v}} = R_z(\vartheta) R_x(I) R_z(\gamma) \hat{\bm{y}} \ ,
\end{equation}
where $R_x$ and $R_z$ are rotation matrices, and we have introduced the unit vectors $\hat{\bm{x}} = (1, 0, 0)$, $\hat{\bm{y}} = (0, 1, 0)$ and $\hat{\bm{z}} = (0, 0, 1)$. 
In the case of an equal-mass inner binary, the quadrupolar magnetic Hamiltonian vanishes.

It is now straightforward to compute the inner orbit-averaged electric and magnetic tidal Hamiltonians, whose full expressions are provided in Appendix~\ref{Appendix A: Tidal moments}. The complete interaction Hamiltonian is
\begin{equation}
\label{Copleteinteracitonhamiltonian}
    \langle H_{\text{tidal}}\rangle=\langle H_{\mathscr{E}}\rangle+\langle H_{\mathscr{B}}\rangle \,.
\end{equation}
It is important to emphasize that the two contributions to the tidal Hamiltonian enter at different PN orders: the electric term appears at 0PN order, while the magnetic term contributes at 0.5PN order.\footnote{The inclusion of the magnetic interaction, which couples the CoM to the relative motion, is consistent within the quadrupolar order in the small-tide approximation, as reviewed in Ref.~\cite{Cocco:2025udb}.}

We can now write the equations of motion for the inner orbital dynamics, corresponding to the evolution of the Delaunay variables. Their evolution can be straightforwardly derived within Hamilton–Jacobi theory: they are governed by a set of first-order differential equations well known in celestial mechanics, namely the Lagrange Planetary Equations (LPE). In particular, in the generalized Lagrange gauge, the LPE for the Delaunay variables form a symplectic system~\cite{Efroimsky:2003pg, Efroimsky:2003mz}. This system reads\footnote{Note that, in the circular limit, the LPE written in terms of the Keplerian orbital elements may exhibit a singular behavior, which can be overcome using a different coordinate system (for details see Refs.~\cite{Deme:2024cvm,Will:2016pgm,Tucker:2021mvo}).}
\begin{equation}
\label{LPE}
\begin{aligned}
    \frac{da}{d\hat{t}} &= \frac{1}{u^{\hat{t}}} \left( -\sqrt{\frac{4a}{GM}} \frac{\partial \mathscr{H}}{\partial \beta}\right) ,\\
    \frac{de}{d\hat{t}} &= \frac{1}{u^{\hat{t}}} \left( \sqrt{\frac{1-e^2}{GMa e^2}} \frac{\partial \mathscr{H}}{\partial \gamma} - \frac{1-e^2}{\sqrt{GMa}e} \frac{\partial \mathscr{H}}{\partial \beta}  \right) , \\
    \frac{dI}{d\hat{t}} &= \frac{1}{u^{\hat{t}}} \frac{1}{\sqrt{GMa(1-e^2)} \sin I} \left( \frac{\partial \mathscr{H}}{\partial \vartheta} - \cos I \frac{\partial \mathscr{H}}{\partial \gamma} \right) , \\
    \frac{d\beta}{d\hat{t}} &= \frac{1}{u^{\hat{t}}} \left( \sqrt{\frac{4a}{GM}} \frac{\partial \mathscr{H}}{\partial a} + \frac{1-e^2}{\sqrt{GMa}e} \frac{\partial \mathscr{H}}{\partial e} \right) , \\
    \frac{d\gamma}{d\hat{t}} &= \frac{1}{u^{\hat{t}}} \left( -\sqrt{\frac{1-e^2}{GMa e^2}} \frac{\partial \mathscr{H}}{\partial e} + \frac{\cot I}{\sqrt{GMa(1-e^2)}} \frac{\partial \mathscr{H}}{\partial I} \right) , \\
    \frac{d\vartheta}{d\hat{t}} &= \frac{1}{u^{\hat{t}}} \left( -\frac{1}{\sqrt{GMa(1-e^2)} \sin I} \frac{\partial \mathscr{H}}{\partial I} \right) ,
\end{aligned}
\end{equation}
where $\mathscr{H}=(\langle H_{\text{inn}} \rangle+\langle H_{\text{tidal}}\rangle)/\mu$. The quantity $u^{\hat{t}}=d\hat{t}/d\hat{\tau}$ represents the gravitational redshift factor, which enables the evolution of the inner binary to be conveniently expressed in terms of the asymptotic time coordinate $\hat{t}$; an explicit expression is provided in Eq.~\eqref{Kerrgeodesicsa}. 

To account for gravitational RR, we supplement these equations with two additional contributions governing the orbital parameters $a$ and $e$. In fact, the leading dissipative contributions enter the dynamics as 2.5PN corrections to the LPE for $a$ and $e$~\cite{Peters:1963ux, Peters:1964zz}
\begin{equation}
\label{Petersequations}
\begin{aligned}
    \left. \frac{da}{d\hat{t}} \right|_{\text{GW}} \!\!&= \!-\frac{1}{u^{\hat{t}}}\frac{64}{5} \frac{G^3 \mu M^2}{c^5 a^3} \frac{1}{(1-e^2)^{7/2}} \!\left( 1 + \frac{73}{24}e^2 + \frac{37}{96}e^4 \right) \!,\\
    \left. \frac{de}{d\hat{t}} \right|_{\text{GW}} \!\!&=\! -\frac{1}{u^{\hat{t}}}\frac{304}{15} \frac{G^3 \mu M^2}{c^5 a^4} \frac{e}{(1-e^2)^{5/2}} \!\left( 1 + \frac{121}{304}e^2 \right) \!.
\end{aligned}
\end{equation}
These equations account for the loss of orbital energy and angular momentum in a weak-field regime, describing the averaged evolution of a binary system due to GW emission. This process decreases the semi-major axis and gradually circularizes the orbit, thereby accelerating periastron precession. Therefore, the inclusion of gravitational back-reaction is crucial in our analysis as it provides a natural mechanism for the system to evolve, and thus go through the various resonances~\cite{Kuntz:2021hhm,Cocco:2025adu, Cocco:2025udb}.


\subsection{\label{sec:Point particle Hamiltonian}Point particle Hamiltonian}

In Sec.~\ref{Sec:Assumptions}, we stated two fundamental assumptions: (i) we neglect RR from the outer orbit; (ii) we treat the tidal interaction within the quadrupolar approximation, which, up to 0.5PN order, allows for a consistent decoupling between the CoM motion and the relative dynamics. With these assumptions, to leading order, the CoM of the inner binary follows a geodesic in Kerr spacetime throughout the entire evolution of the system.
The corresponding Hamiltonian is given by 
\begin{equation}
\label{Pointparticlehamiltonian}
    H_{\text{pp}}=\frac{1}{2}g^{\mu\nu} p_{\mu} p_{\nu}
\end{equation}
where $g_{\mu\nu}$ is the Kerr metric given in Eq.~\eqref{Kerrmetric}.

The associated equations of motion are the Kerr geodesic equations, which can be written with respect to the proper time as
\begin{subequations}
\label{Kerrgeodesics}
\begin{align}
    \label{Kerrgeodesicsa}
    &\Sigma\frac{d\hat{t}}{d\hat{\tau}} \!=\! \frac{\hat{r}^2 + \mathrm{a}^2}{\Delta}\! \left( \hat{E}(\hat{r}^2 + \mathrm{a}^2)\! -\! \mathrm{a}\hat{L}_{\hat{z}} \right)\! - \!\mathrm{a}^2 \hat{E}(1 - \hat{z}^2)\! +\! \mathrm{a}\hat{L}_{\hat{z}} \,,\\
    \label{Kerrgeodesicsb}
    &\Sigma^2\!\left( \!\frac{1}{c}\frac{d\hat{r}}{d\hat{\tau}} \right)^{\!\!\!2} \!\!=\!\left( \hat{E}(\hat{r}^2 + \mathrm{a}^2) - \mathrm{a}\hat{L}_{\hat{z}} \right)^2 - \!\!\Delta \left( \hat{r}^2 + \hat{K} \right),\quad \\
    \label{Kerrgeodesicsc}
    &\Sigma^2\!\left( \!\frac{1}{c}\frac{d\hat{z}}{d\hat{\tau}} \right)^{\!\!\!2} \!\!=\!  (\hat{K}\!-\!\mathrm{a}^2\hat{z}^2)(1\!-\!\hat{z}^2)\!-\!\left(\hat{L}_{\hat{z}}\!-\!\mathrm{a}(1\!-\!\hat{z}^2)\hat{E}\right)^{\!\!2},\\
    \label{Kerrgeodesicsd}
    &\Sigma \frac{1}{c}\frac{d\hat{\phi}}{d\hat{\tau}} = \frac{\mathrm{a}}{\Delta} \left( \hat{E}(\hat{r}^2 + \mathrm{a}^2) - \mathrm{a}\hat{L}_{\hat{z}} \right) + \frac{\hat{L}_{\hat{z}}}{1 - \hat{z}^2} - \mathrm{a}\hat{E} \,.
\end{align}
\end{subequations}
Following Refs.~\cite{Warburton:2010eq,Warburton:2013yj}, we adopt the convention in which $\hat{L}_{\hat{z}}$ is positive and distinguish \textit{prograde} and \textit{retrograde} orbits by the sign of the spin parameter, $\rm a>0$ and $\rm a<0$, respectively.\footnote{Prograde and retrograde orbits can also sometimes be referred to as \textit{co-rotating} and \textit{counter-rotating} configurations.}
In what follows, we focus on bound geodesics, for which it is convenient to map the constants of motion into the set of orbital elements
\begin{equation} 
\label{setouterorbitalparameters}
    (\hat{E},\hat{L}_{\hat{z}},\hat{{K}})\mapsto(\hat{p}, \hat{e}, \hat{z}_1)~,
\end{equation}
namely the semi-latus rectum $\hat{p}=\hat{a}(1-\hat{e}^2)$, the eccentricity $\hat{e}$ and the polar turning point $\hat{z}_1$ (root of Eq.~\eqref{Kerrgeodesicsc}).
Two other roots, $\hat{r}_{3,4}$, can be found for Eq.~\eqref{Kerrgeodesicsb} and one more, $\hat{z}_2$, for Eq.~\eqref{Kerrgeodesicsc}. Compact and operational expressions for all the newly introduced quantities can be found in Refs.~\cite{Schmidt:2002qk, vandeMeent:2019cam}.

We briefly review two approaches for solving these equations: the first is optimized for full numerical integration, while the second is better suited for analytical manipulations. We employ both in our semi-analytical framework.


\subsubsection*{Radial and polar variables}

To obtain numerical solutions of the geodesic equations~\eqref{Kerrgeodesicsa}--\eqref{Kerrgeodesicsd}, we adopt standard parametrizations of the radial $\hat{r}$ and polar $\hat{z}$ coordinates, as introduced, for example, in Refs.~\cite{1983mtbh.book.....C} and~\cite{Drasco:2003ky}. In particular, we introduce the relativistic radial and polar anomalies $\hat{\psi}_{\hat{r}},\hat{\psi}_{\hat{z}}\in[0,2\pi]$ such that\footnote{Note that our $\hat{\psi}_{\hat{r}}$ coincides with the one introduced in Ref.~\cite{1983mtbh.book.....C}and our $\hat{\psi}_{\hat{z}}$ coincides with the angle $\chi$ of Ref.~\cite{Drasco:2003ky}.}
\begin{equation}
\label{Radialpolararametrization}
    \hat{r}=\frac{\hat{a}(1-\hat{e}^2)}{1+\hat{e}\cos{\hat{\psi}_{\hat{r}}}}\,,\quad \hat{z}=\hat{z}_1\cos{\hat{\psi}_{\hat{z}}}~.
\end{equation}
Substituting this parametrization into the geodesic equations~\eqref{Kerrgeodesicsb}--\eqref{Kerrgeodesicsc} directly yields the evolution equations for $\hat{\psi}_{\hat{r}},\hat{\psi}_{\hat{z}}$. We use the parametrization~\eqref{Radialpolararametrization} also for the evolution of Marck's angle $\hat{\Psi}$. Finally, the geodesic equation~\eqref{Kerrgeodesicsa} can be rearranged to obtain an expression for the gravitational redshift factor $u^{\hat{t}} = d\hat{t}/d\hat{\tau}$, which allows us to describe the evolution of the triple system in terms of the asymptotic time $\hat{t}$, the natural time coordinate for a distant observer far from the SMBH. This set of equations fully specifies the outer motion and, being expressed in terms of periodic variables, is well suited for numerical integration.


\subsubsection*{Action-angle variables}

One of the main difficulties in trying to obtain analytic solutions to Eqs.~\eqref{Kerrgeodesicsa}--\eqref{Kerrgeodesicsd} arises from the fact that Eqs.~\eqref{Kerrgeodesicsb} and~\eqref{Kerrgeodesicsc} couple the radial and polar degrees of freedom. Nevertheless, in both equations, this coupling enters solely through the function $\Sigma$. This observation naturally motivates the introduction of the \textit{Mino(-Carter) time} $\hat{\lambda}$~\cite{Mino:2003yg}, defined as
\begin{equation}
\label{Minotime}
    d\hat{\tau}=\Sigma \,d\hat{\lambda}\,.
\end{equation}
With this choice of time parametrization, closed-form solutions of the geodesic equations and Eq.~\eqref{Marckevolution} for the evolution of $\hat{\Psi}$ can be obtained by employing an \textit{action-angle variable} formalism, in which the dynamics is described in terms of conserved actions and linearly evolving angles (or \textit{phases})
\begin{equation}
\label{Anglevariablesmino}
    q_{A}(\hat{\lambda}) = \Upsilon_{A}\hat{\lambda}\,, \quad A = (\hat{t}, \hat{r}, \hat{z}, \hat{\phi},\hat{\Psi})\,,
\end{equation}
where $\Upsilon_{A}$ denote the corresponding frequencies with respect to the Mino time.
In this framework, the geodesic equations and Marck's equation can be solved as functions of the Mino time in closed form using elliptic functions~\cite{Schmidt:2002qk,Drasco:2003ky,Fujita:2009bp,vandeMeent:2019cam}. Following the notation of Ref.~\cite{Cocco:2025udb}, the solutions take the implicit form
\begin{equation}
\label{Schematicsolution}
\begin{aligned}
    \hat{t}(\hat{\lambda})&=q_{\hat{t}}(\hat{\lambda})+\varphi_{\hat{t}\hat{r}}(q_{\hat{r}}(\hat{\lambda}))+\varphi_{\hat{t}\hat{z}}(q_{\hat{z}}(\hat{\lambda}))\,,\\
    \hat{r}(\hat{\lambda})&=\varphi_{\hat{r}\hat{r}}(q_{\hat{r}}(\hat{\lambda}))\,,\\
    \hat{z}(\hat{\lambda})&=\varphi_{\hat{z}\hat{z}}(q_{\hat{z}}(\hat{\lambda}))\,,\\
    \hat{\phi}(\hat{\lambda})&=q_{\hat{\phi}}(\hat{\lambda})+\varphi_{\hat{\phi}\hat{r}}(q_{\hat{r}}(\hat{\lambda}))+\varphi_{\hat{\phi}\hat{z}}(q_{\hat{z}}(\hat{\lambda}))\,,\\
    \hat{\Psi}(\hat{\lambda})&=q_{\hat{\Psi}}(\hat{\lambda})+\varphi_{\hat{\Psi}\hat{r}}(q_{\hat{r}}(\hat{\lambda}))+\varphi_{\hat{\Psi}\hat{z}}(q_{\hat{z}}(\hat{\lambda}))\,,
\end{aligned}
\end{equation}
which highlights the decoupling of the radial and polar degrees of freedom. The phases $q_{\hat{t}}$ and $q_{\hat{\Psi}}$ account for the secular accumulation of $\hat{t}$ and $\hat{\Psi}$, as expected from circular, equatorial orbits; $\varphi_{A B}$, with $A,B=(\hat{t}, \hat{r}, \hat{z},  \hat{\phi},\hat{\Psi})$, are elliptic functions of $q_{\hat{r}}$ and $q_{\hat{z}}$ accounting for the oscillatory dynamics associated to the radial and polar degrees of freedom~\cite{Fujita:2009bp}. Ref.~\cite{vandeMeent:2019cam} provides compact analytical expressions for all these functions, as well as for the fundamental frequencies $\Upsilon_A$.

We will ultimately express our results in terms of the asymptotic time $\hat{t}$. In this parametrization, the phases are defined as
\begin{equation}
\label{Anglevariablesasy}
    Q_{A}(\hat{t}) = \Omega_{A}\hat{t}\,, \quad A = (\hat{r}, \hat{z}, \hat{\phi}, \hat{\Psi})\,,
\end{equation}
where the quantities $\Omega_A$ are the fundamental frequencies of the outer motion in coordinate time. These are related to the Mino-time frequencies $\Upsilon_A$ through a simple rescaling
\begin{equation}
\label{Switchfrequencies}
    \Omega_A=\frac{\Upsilon_A}{\Upsilon_{\hat{t}}}\,,\quad A = (\hat{r}, \hat{z},  \hat{\phi},\hat{\Psi})\,.
\end{equation}
Explicit expressions for these frequencies will be provided in the next section.
Note that the azimuthal angle variable $\hat{\phi}$ does not appear in the tidal Hamiltonian (see Eqs. (\ref{Fullelectrichamiltonian})-(\ref{Fullmagnetichamiltonian})) and can therefore be neglected in the evolution, as already emphasized in Refs.~\cite{Cocco:2025adu, Cocco:2025udb}.


\section{\label{Sec:Fourier}Perturbative Fourier analysis of precession resonances}

In this section, we focus on quasi-circular and quasi-equatorial orbits in Kerr spacetime. 
To this aim, we perform a perturbative analysis of the quantities introduced in Sec.~\ref{sec:Point particle Hamiltonian} for small values of the outer eccentricity $\hat{e}$ and of the polar turning point $\hat{z}_{1}$ defined in Eq.~\eqref{Radialpolararametrization}.
This procedure yields compact and analytically tractable expressions, which can be used to carry out a Fourier analysis of the interaction between the inner binary and the SMBH, thereby revealing the presence of precession resonances.


\subsection{\label{Sec: Precession resonances in Kerr spacetime}Fourier expansion of the tidal Hamiltonian}

Precession resonances occur when the precession frequency of the inner binary, evolving according to Eq.~\eqref{LPE}, becomes commensurate with a linear combination of the fundamental frequencies of the outer orbital motion, $\Upsilon_A$ with $A = (\hat{r}, \hat{z}, \hat{\Psi})$ (or $\Omega_A$ when expressed in asymptotic time). 
Expressed in terms of the asymptotic-time variables, this condition can be written as
\begin{align}
\label{Resonantcondition2}
    q\frac{d\gamma}{d\hat{t}}&=k \Omega_{\hat{r}}+l \Omega_{\hat{z}}+m \Omega_{\hat{\Psi}}\,,
\end{align}
where $q$, $k$, $l$ and $m$ are integers. 
From a Hamiltonian perspective, resonances require a dependence on combinations of the inner angle $\gamma$ and the outer phases. 
Such a dependence is provided only by the quadrupolar interaction Hamiltonian $\langle H_{\text{tidal}} \rangle$, which encodes the coupling between the inner binary and the SMBH through the tidal moments. Neither $\langle H_{\text{inn}} \rangle$ nor $H_{\text{pp}}$ depends on $\gamma$ together with the outer phases.
The redshift factor $u^{\hat t}$, entering the generator $\langle H\rangle/u^{\hat t}$ of the evolution in asymptotic time, does not depend on $\gamma$ and therefore cannot generate resonant terms by itself.
However, since it depends on the outer phases, it modulates the Fourier modes of the tidal Hamiltonian, and we consistently include it.

To analytically investigate these resonances within a perturbative framework, a necessary step is to express the electric and magnetic tidal moments appearing in Eq.~\eqref{Copleteinteracitonhamiltonian} in terms of angle variables by substituting $\hat{r}$, $\hat{z}$, and $\hat{\Psi}$ with their corresponding expressions in terms of elliptic functions. 
In the case of a Schwarzschild background, this procedure can be easily carried out directly using the asymptotic time; in contrast, in the Kerr case considered here, as highlighted in Sec.~\ref{Sec: Perturbative approach}, it is more convenient to work with the Mino-time solutions of the geodesic equations~\eqref{Schematicsolution}. 
We are however interested in expressing the Fourier coefficients in asymptotic-time variables. To do this, we follow below the prescription outlined in Ref.~\cite{Drasco:2003ky} to obtain the Fourier coefficients in asymptotic time.

First, we schematically decompose and Fourier-expand the generators of the evolution in the asymptotic time, $\langle H_{\mathscr{E}}\rangle/u^{\hat t}$ and $\langle H_{\mathscr{B}}\rangle/u^{\hat t}$, with $\langle H_{\mathscr{E}}\rangle$ and $\langle H_{\mathscr{B}}\rangle$ given in Eq.~\eqref{Copleteinteracitonhamiltonian}, as
\begin{equation}
\label{Fourierexpansion}
\begin{aligned}
     \frac{\langle H_{\mathscr{E}}\rangle}{u^{\hat t}}=\frac{1}{2}c^2\sum_q h^{\mathscr{E},\text{inn}}_q\, e^{iq\gamma}\sum_{k,l,m}h^{\mathscr{E},\text{out}}_{klm}e^{-i(k\Omega_{\hat{r}}+l\Omega_{\hat{z}}+m\Omega_{\hat{\Psi}})\hat{t}}\,,\\
    \frac{\langle H_{\mathscr{B}}\rangle}{u^{\hat t}}=\frac{2}{3}c\sum_q h^{\mathscr{B},\text{inn}}_q\, e^{iq\gamma}\sum_{k,l,m}h^{\mathscr{B},\text{out}}_{klm}e^{-i(k\Omega_{\hat{r}}+l\Omega_{\hat{z}}+m\Omega_{\hat{\Psi}})\hat{t}}\,.
\end{aligned}
\end{equation}
For both the electric and magnetic tidal Hamiltonians, the first summation corresponds to the direct Fourier expansion of the inner dynamics, as encoded in $\mathscr{Q}^{ij}$, defined in Eq.~\eqref{Averagedelectricandmagnetictidalhamiltonian1}, and $\mathscr{T}^{ij}$, defined in Eq.~\eqref{Averagedelectricandmagnetictidalhamiltonian2}.
In particular, since $\hat{\bm u}$ and $\hat{\bm v}$ depend on $\gamma$ only through $e^{\pm i\gamma}$, $\langle\mathscr{Q}^{ij}\rangle$, being quadratic in $(\hat{\bm u},\hat{\bm v})$, restricts the inner harmonics to $q=0,\pm2$, whereas $\langle\mathscr{T}^{ij}\rangle\propto u^i(\hat{\bm u}\times\hat{\bm v})^j$, being linear in $\hat{\bm u}$, allows only $q=\pm1$.\footnote{This naturally agrees with Refs.~\cite{Cocco:2025adu} and~\cite{Cocco:2025udb} despite the different background spacetime, since the corresponding tidal tensors couple to the same tensors encoding the inner binary dynamics, $\mathscr{Q}^{ij}$ and $\mathscr{T}^{ij}$.}
Only the harmonics with $q\neq0$ can give rise to resonances. However, although the $q=0$ electric terms are independent of $\gamma$ and therefore non-resonant, they must be retained as they contain the secular contribution $(k,l,m)=(0,0,0)$, which drives the tidal precession of $\gamma$ and $\vartheta$. No such term arises in the magnetic sector, \textit{i.e.}, $h^{\mathscr{B},\text{inn}}_0=0$.

The second summation in Eq.~\eqref{Fourierexpansion} instead captures the contribution from the outer motion and corresponds to the Fourier expansion of the electric and magnetic tidal moments, given in Eqs.~\eqref{Explicitelectrictidal} and~\eqref{Explicitmagnetictidal}, divided by the redshift factor ${u^{\hat t}}$ and evaluated along the geodesic~\eqref{Schematicsolution}.
Since the closed-form solutions are given in Mino time, while the expansion~\eqref{Fourierexpansion} is in asymptotic time, we obtain the coefficients by generalizing the prescription of Ref.~\cite{Drasco:2003ky}.
We first split the Mino-time solution for $\hat t$ of Eq.~\eqref{Schematicsolution} into a secular and an oscillating
part,
\begin{align}
\label{Drascoprescription1}
    \hat{t}(\hat{\lambda})&=\Upsilon_{\hat{t}}\hat{\lambda}+\Delta \hat{t}(\hat{\lambda})\,,\\
    \frac{d\hat{t}(\hat{\lambda})}{d\hat{\lambda}}&=\Upsilon_{\hat{t}}+\frac{d(\Delta \hat{t}(\hat{\lambda}))}{d\hat{\lambda}}\,,
\end{align}
with $\Delta\hat{t}=\varphi_{\hat{t}\hat{r}}(q_{\hat{r}})+\varphi_{\hat{t}\hat{z}}(q_{\hat{z}})$ and, for a generic frequency $\omega$, define the electric and magnetic functionals as\footnote{This notation is rather schematic, as each component of the two tensors must be evaluated individually. Accordingly, the two functionals should carry two spatial indices, which we drop here for the sake of notational simplicity.}
\begin{equation}
\label{Drascoprescription2}    
\begin{aligned}         
    \mathcal{F}^{\mathscr{E}}&\!=\!\frac{d\hat{t}(\hat{\lambda})}{d\hat{\lambda}}e^{i\omega \Delta\hat{t}(\hat{\lambda})}\frac{\mathscr{E}}{u^{\hat t}}\!=\!\!\sum_{k,l,m}\!\!\mathcal{F}^{\mathscr{E}}_{klm}(\omega)e^{-i(k\Upsilon_{\hat{r}}+l\Upsilon_{\hat{z}}+m\Upsilon_{\hat{\Psi}})\hat{\lambda}},\\
    \mathcal{F}^{\mathscr{B}}&\!=\!\frac{d\hat{t}(\hat{\lambda})}{d\hat{\lambda}}e^{i\omega \Delta\hat{t}(\hat{\lambda})}\frac{\mathscr{B}}{u^{\hat t}}\!=\!\!\sum_{k,l,m}\!\!\mathcal{F}^{\mathscr{B}}_{klm}(\omega)e^{-i(k\Upsilon_{\hat{r}}+l\Upsilon_{\hat{z}}+m\Upsilon_{\hat{\Psi}})\hat{\lambda}},
\end{aligned}
\end{equation}
where $\mathscr{E}$, $\mathscr{B}$ and $u^{\hat t}$ are expressed in terms of the solutions of the geodesic equations given in Eq.~\eqref{Schematicsolution}. Since the tidal moments depend on the Marck's angle $\hat{\Psi}=q_{\hat{\Psi}}+\varphi_{\hat{\Psi}\hat{r}}(q_{\hat{r}})+\varphi_{\hat{\Psi}\hat{z}}(q_{\hat{z}})$, the functionals in Eq.~\eqref{Drascoprescription2} are $2\pi$-periodic functions on the three-torus $(q_{\hat{r}},q_{\hat{z}},q_{\hat{\Psi}})$; this generalizes the prescription of Ref.~\cite{Drasco:2003ky}, which considers functions of $\hat{r}$ and $\hat{z}$ only. The modes $\mathcal{F}^{\mathscr{E},\mathscr{B}}_{klm}$ can then be derived as Fourier coefficients of the functionals $\mathcal{F}^{\mathscr{E},\mathscr{B}}$ in the Mino time representation. Then the asymptotic time Fourier coefficients $h_{klm}^{\mathscr{E},\text{out}}$ and $h_{klm}^{\mathscr{B},\text{out}}$ in Eq.~\eqref{Fourierexpansion} can be computed as~\cite{Drasco:2003ky} 
\begin{equation}
\label{Coefficientsouter}
\begin{aligned}
    h_{klm}^{\mathscr{E},\text{out}}=\frac{\mathcal{F}^{\mathscr{E}}_{klm}(k\Omega_{\hat{r}}+l \Omega_{\hat{z}}+m\Omega_{\hat{\Psi}})}{\Upsilon_{\hat{t}}}\,,\\
    h_{klm}^{\mathscr{B},\text{out}}=\frac{\mathcal{F}^{\mathscr{B}}_{klm}(k\Omega_{\hat{r}}+l \Omega_{\hat{z}}+m\Omega_{\hat{\Psi}})}{\Upsilon_{\hat{t}}}\,.
\end{aligned}
\end{equation}
Combining the inner and outer expansions and using the reality of the Hamiltonians, we can write
\begin{equation}
\label{Fourierexpansionfinal}
\begin{aligned}
    \frac{\langle H_{\mathscr{E}}\rangle}{u^{\hat t}}&=\sum_{k,l,m}h^{\mathscr{E}}_{0klm}e^{-i(k\Omega_{\hat{r}}+l\Omega_{\hat{z}}+m\Omega_{\hat{\Psi}})\hat{t}}+\\
    &\,\,\,\,\,\,\,\sum_{k,l,m}2\mathrm{Re}\left[h^{\mathscr{E}}_{2klm}e^{i 2\gamma-i(k\Omega_{\hat{r}}+l\Omega_{\hat{z}}+m\Omega_{\hat{\Psi}})\hat{t}}\right]\,,\\
    \frac{\langle H_{\mathscr{B}}\rangle}{u^{\hat t}}&=\sum_{k,l,m}2\mathrm{Re}\left[h^{\mathscr{B}}_{1klm}e^{i \gamma-i(k\Omega_{\hat{r}}+l\Omega_{\hat{z}}+m\Omega_{\hat{\Psi}})\hat{t}}\right]\,,
\end{aligned}
\end{equation}
where the sums run over all integers $(k,l,m)$, $h^{\mathscr{E}}_{qklm}=\frac{1}{2}c^2\,h^{\mathscr{E},\text{inn}}_qh^{\mathscr{E},\text{out}}_{klm}$ and $h^{\mathscr{B}}_{qklm}=\frac{2}{3}c\,h^{\mathscr{B},\text{inn}}_qh^{\mathscr{B},\text{out}}_{klm}$.

In this work, we retain all the possible Fourier modes arising from the projection procedure, so that the tidal Hamiltonian is fully reconstructed at the perturbative order we consider. 
Since by construction $J_\gamma > 0$, the 1PN Hamiltonian gives $\dot\gamma>0$ and the system can only encounter resonances with $k\Omega_{\hat{r}}+l\Omega_{\hat{z}}+m\Omega_{\hat{\Psi}}>0$.
Which combinations satisfy this condition depends on the interplay between the outer frequencies or, equivalently, on the outer orbital elements.
Each mode $(k,l,m)$ comes with its partner $(-k,-l,-m)$, whose frequency combination has the opposite sign and describes the same resonance in the inner binary but with reversed orbital motion.
This reflects the symmetries of $H_{\mathscr{E}}$ and $H_{\mathscr{B}}$ under inversion of the inner motion, $\gamma\to\pi-\gamma$, $\vartheta\to\pi+\vartheta$, and $I\to\pi-I$~\cite{Liu_2014}. Specifically, $H_{\mathscr{E}}$ is symmetric under this transformation, whereas $H_{\mathscr{B}}$ is antisymmetric. 
Consistently, inserting this transformation in Eq.~\eqref{Fourierexpansionfinal} and using $\mathrm{Re}\,z=\mathrm{Re}\,z^*$, one finds that the Fourier coefficient associated with $(k,l,m)$ is related to that of $(-k,-l,-m)$ via
\begin{equation}
\label{Symmetrycoefficients}
    h^{\mathscr{E},\mathscr{B}}_{q~-k~-l~-m}(I,\vartheta)=\Big[h^{\mathscr{E},\mathscr{B}}_{qklm}(\pi-I,\pi+\vartheta)\Big]^*\,,
\end{equation}
with the same sign for the electric ($q=2$) and the magnetic ($q=1$) contributions. In the latter, the antisymmetry of $H_{\mathscr{B}}$ is compensated by the factor $e^{i\pi}=-1$ arising from $e^{i\gamma}\to e^{i(\pi-\gamma)}$.


\subsection{\label{Sec: Perturbative approach}Quasi-circular and quasi-equatorial regime}
The coupling between the outer motion and the internal dynamics is encoded in the interaction Hamiltonian given in Eq.~\eqref{Copleteinteracitonhamiltonian}.
Specifically, in order to use the prescription outlined in the previous section, one must write the tidal moments using Eq.~\eqref{Schematicsolution}.
However, since the expressions in Eq.~\eqref{Schematicsolution} are given in terms of elliptic functions, in their exact form they are difficult to work with for deriving analytic expressions for the Fourier coefficients.
We therefore follow Refs.~\cite{Cocco:2025adu,Cocco:2025udb} by considering a perturbative expansion in the eccentricity of the outer orbit ($\hat{e} \ll 1$) and, as a novelty for the Kerr scenario, also a perturbative expansion in the inclination angle of the outer orbit ($\hat{z}_1 \ll 1$). 
For the functions $\varphi_{AB}$ and the frequencies $\Omega_A$, we perform our computation neglecting terms of order  $\mathscr{O}(\hat{e}^2)$ and $\mathscr{O}(\hat{z}_1^2)$.
As shown in Appendix~\ref{Appendix C: Results of the Double Expansion}, the expansion of the outer motion does not contain terms of $\mathscr{O}(\hat{e}\,\hat{z}_1)$. Such terms, however, naturally arise in the tidal Hamiltonian. As discussed below, we retain the terms of order $\mathscr{O}(\hat{e}\,\hat{z}_1)$, since they provide the leading-order resonances arising from the coupling of the radial and polar degrees of freedom, even though terms of $\mathscr{O}(\hat{e}^2)$ and $\mathscr{O}(\hat{z}_1^2)$ could be of comparable magnitude.
This is consistent since we want to retain only the first-order correction for each class of resonances, as it will be clarified in Sec.~\ref{Sec: Resonances classification}.

We conveniently introduce the dimensionless quantities
\begin{equation}
\label{Dimensionlessquantities}
    \hat{\sigma}=\hat{a}\frac{c^2}{G M_*}\,,\quad \chi=\mathrm{a}\frac{c^2}{G M_*}\,.
\end{equation}
The first parameter remains finite in the strong-field regime but diverges in the weak-field limit; accordingly, we recover the Newtonian results as $\hat{\sigma} \to \infty$. The second is the \textit{dimensionless spin parameter}, which allows us to validate our results by comparison with Refs.~\cite{Cocco:2025adu,Cocco:2025udb} in the limit $\chi \to 0$. 
With the convention adopted in this paper, the parameter $\chi$ provides a natural way to distinguish prograde ($\chi \in ]0,1]$) and retrograde ($\chi \in [-1,0[$) configurations of the inner binary around the supermassive Kerr black hole. 
Note that, while performing the double expansion in the small orbital parameters $\hat{e}$ and $\hat{z}_1$, no assumption is made on the spin parameter $\chi$. This allows for arbitrarily high values of the spin parameter, corresponding to near-extremal BHs.

In Appendix~\ref{Appendix C: Results of the Double Expansion}, we present the results of the double expansion of the elliptic functions and of the relevant frequencies within the action-angle formalism, as well as of the constants of motion in Kerr spacetime. A consistency check consists in verifying that our frequencies $\Omega_A$ and constants of motion reduce to those of Ref.~\cite{Cocco:2025udb} in the limit $\chi \to 0$, while the Newtonian results are recovered in the weak-field limit $\hat{\sigma} \to \infty$.

Conversely, we cannot compare the solutions $\varphi_{AB}$, given in Eqs.~\eqref{Schematicsolution}, directly to the functions $f_{AB}$ provided in Refs.~\cite{Cocco:2025adu,Cocco:2025udb}. This is because $\varphi_{AB}$ are expressed in terms of the Mino-time, while $f_{AB}$ are in terms of the asymptotic time.\footnote{One should perform a reparametrization of the time variable to obtain the expression in the asymptotic time variables of our elliptic functions $\varphi_{AB}$.}


\subsection{\label{Sec: Resonances classification}Classification of resonances}

Using the procedure outlined in
Sec.~\ref{Sec: Precession resonances in Kerr spacetime} and the perturbative framework introduced in Sec.~\ref{Sec: Perturbative approach}, we identify the precession resonances that can arise in Kerr spacetime from the non-vanishing Fourier coefficients
$h_{qklm}^{\mathscr{E},\mathscr{B}}$ in Eq.~\eqref{Fourierexpansionfinal}.
Near a resonance, the corresponding oscillatory
contribution in Eq.~\eqref{Fourierexpansionfinal}
varies slowly and can accumulate coherently,
leading to significant changes in the orbital
parameters of the inner binary.
As the inner binary shrinks due to RR, its precession frequency increases, allowing the system to pass through successive resonances, as described by Eq.~\eqref{Petersequations}.
The combination of frequencies defining each resonance determines when it can occur during the inspiral, while the corresponding Fourier coefficient determines the strength of the resonant contribution to the orbital dynamics.

Depending on which outer frequencies enter the resonance condition in Eq.~\eqref{Resonantcondition2}, we can group precession resonances in four distinct classes, for both the electric and magnetic sectors:
\begin{itemize}
    \item[(i)] “Circular-equatorial" resonances: these arise independently of both the orbital eccentricity and the polar motion, \textit{i.e.}, they persist in the limit $\hat{e}, \hat{z}_1 \to 0$. While including the full dependence on the spin, we can check that in the limit $\chi \to 0$, our results consistently reduce to those found in Refs.~\cite{Cocco:2025adu, Cocco:2025udb}.
    \item[(ii)] “Radial" resonances: these involve only the radial degree of freedom and 
    the associated perturbative parameter is the outer orbital eccentricity $\hat{e}$. The corresponding Fourier coefficients take the form $h_{qk0m}^{\mathscr{E},\mathscr{B}}$ and their Schwarzschild and Newtonian counterparts can be recovered in the limits $\chi\to0$ and $\hat{\sigma}\rightarrow \infty$, respectively. 
    \item[(iii)] “Polar" resonances: these involve only the polar degree of freedom and the associated perturbative parameter is the polar turning point $\hat{z}_1$. The corresponding Fourier coefficients take the form $h_{q0lm}^{\mathscr{E},\mathscr{B}}$ and, consistently, vanish in the Schwarzschild and Newtonian limit.
    \item[(iv)] “Radial-polar" resonances: these involve both the radial and polar degrees of freedom, and therefore depend on both $\hat{e}$ and $\hat{z}_1$. The corresponding Fourier coefficients take the most general form $h_{qklm}^{\mathscr{E},\mathscr{B}}$ and also vanish in the Schwarzschild and Newtonian limit.
\end{itemize}
This classification is exhaustive and holds at any perturbative order.
The order of the expansion only determines which resonances within each class are captured.
Here, we expand the outer motion independently at linear order in $\hat{e}$ and $\hat{z}_1$, and mixed terms of $\mathscr{O}(\hat{e}\,\hat{z}_1)$ naturally emerge in the tidal Hamiltonian, giving rise to the class (iv). 
For each class, we therefore retain only up to the leading-order contributions, namely $\mathscr{O}(\hat{e})$, $\mathscr{O}(\hat{z}_1)$ and $\mathscr{O}(\hat{e}\,\hat{z}_1)$ for classes (ii), (iii) and (iv), respectively.
We neglect higher-order terms, e.g., of $\mathscr{O}(\hat{e}^2)$ or $\mathscr{O}(\hat{z}_1^2)$, which would yield additional resonances within classes (ii) and (iii).
However, in the case where $\hat{e}$ and $\hat{z}_1$ are of the same order, these could give contributions comparable to those of class (iv). We postpone the analysis of these types of contributions to future work.

We can summarize our results by categorizing the non-vanishing Fourier coefficients into these four classes and according to their electric or magnetic nature. In Table~\ref{Tableresonances}, 0PN (electric) and 0.5PN (magnetic) resonances are compactly identified by dividing the resonance condition in Eq.~\eqref{Resonantcondition2} by $q$,
\begin{equation}
\label{Resonanceconditionsemiintegers}
    \dot{\gamma}=k\Omega_{\hat{r}}+l\Omega_{\hat{z}}+m\Omega_{\hat{\Psi}}\,,\quad k,l,m\in\frac{1}{2}\mathbb{Z}\,,
\end{equation}
where now $(k,l,m)$ remain integers in the magnetic sector ($q=1$), while they can also take half-integer values in the electric one ($q=2$). This labelling is used in Table~\ref{Tableresonances} and in Figs.~\ref{Fig: equatorial_plot}--\ref{Fig: radialpolar_plot}, whereas $(k,l,m)$ are integers in the Fourier coefficients $h_{qklm}^{\mathscr{E},\mathscr{B}}$ $(k,l,m)$.
\begin{table}[t]
\centering
\caption{Resonance conditions in the perturbative scheme discussed in Sec.~\ref{Sec: Resonances classification}.}
\label{Tableresonances}
\vspace{2mm}

\renewcommand{\arraystretch}{1.25}

\begin{tabular}{l|c|c|c|c}
\toprule
\multicolumn{5}{c}{$
\dot{\gamma} = k\Omega_{\hat{r}} + l\Omega_{\hat{z}} + m\Omega_{\hat{\Psi}},
\quad k,l,m \in \frac{1}{2}\mathbb{Z}
$} \\
\midrule
& (i) & (ii) & (iii) & (iv) \\
\midrule

\multirow{4}{*}{0PN}
& $(0,0,1)$
& $(1/2,0,1)$
& $(0,1/2,1/2)$
& $(1/2,1/2,1/2)$ \\

&
& $(-1/2,0,1)$
& $(0,1/2,- 1/2)$
& $(-1/2,1/2,1/2)$ \\

&
& $(1/2,0,0)$
&
& $(1/2,-1/2,1/2)$ \\

&
&
&
& $(1/2,1/2,-1/2)$ \\[1mm]

\midrule

\multirow{6}{*}{0.5PN}
& $(0,0,1)$
& $(1,0,1)$
& $(0,1,2)$
& $(1,1,2)$ \\

&
& $(-1,0,1)$
& $(0,-1,2)$
& $(-1,1,2)$ \\

&
&
& $(0,1,0)$
& $(1,-1,2)$ \\

&
&
&
& $(-1,-1,2)$ \\

&
&
&
& $(1,1,0)$ \\

&
&
&
& $(-1,1,0)$ \\

\bottomrule
\end{tabular}
\end{table}

The explicit expressions for these coefficients are provided in the complementary $\mathsf{Mathematica}$ notebook. Here, a second procedure for computing the Fourier modes is also presented, thereby independently validating our results.
We have verified that our results satisfy the correct Newtonian limit given in Ref.~\cite{Kuntz:2021hhm} and, up to the inclusion of the redshift factor in Eq.~\eqref{Fourierexpansionfinal}, the Schwarzschild limit given in Refs.~\cite{Cocco:2025adu,Cocco:2025udb}.

It is clear that the resonance condition~\eqref{Resonanceconditionsemiintegers} derived in the Kerr geometry admits a richer spectrum of resonances compared to the Schwarzschild case. The polar degree of freedom, which is trivial in Schwarzschild spacetime due to spherical symmetry, now gives rise to purely polar resonances in the electric and magnetic sectors. Furthermore, the interplay between the radial and polar degrees of freedom leads to many additional new resonances in the two sectors.

Finally, from Table~\ref{Tableresonances} one can notice a particularly interesting subset of the resonances, present in both sectors, which involves combinations of the outer frequencies such that the right-hand side of Eq.~\eqref{Resonanceconditionsemiintegers} vanishes in the Newtonian limit.
This cancellation corresponds to precessional motions that are absent in Newtonian gravity. 
A strong-gravity description of the outer motion, breaking the degeneracy of the fundamental frequencies, gives rise to outer precession frequencies that can become commensurate with the inner 1PN periastron precession frequency~\cite{Schmidt:2002qk}.
These kinds of resonances thus couple relativistic precessions of the two orbits, rather than the inner precession with the outer orbital motion itself. 
For instance, the $(0,1/2,-1/2)$ condition arises from the precession of the outer orbital plane.


\section{\label{Sec:Numerical}Numerical analysis}

In this section, we solve numerically the equations for the evolution of the inner binary's orbital quantities, namely Eqs.~\eqref{LPE} and~\eqref{Petersequations}, coupled to the outer motion. The system is evolved through the Hamiltonian~\eqref{decompositionHamiltonian}, averaged over the inner motion.
Numerical values of the orbital parameters, masses and spin are chosen to satisfy the model's assumptions outlined in Sec.~\ref{Sec:Assumptions}.
This approach is valid for finite values of the outer parameters $\hat{e}$ and $\hat{z}_1$.
However, to validate our analytical results derived in Sec.~\ref{Sec: Resonances classification}, we retain perturbative values for $\hat{e}$ and $\hat{z}_1$. This is consistent with the perturbative scheme underlying our model. Other approaches can be used to probe outer orbital dynamics characterized by high eccentricities~\cite{10.1093/mnras/staa2492, Ginat:2025kch}.

We restrict the numerical analysis to inner orbits precessing with positive precession frequencies. Therefore, only certain resonant combinations will be matched for every class of resonances (i)-(iv), depending on the specific choice of masses, spin and orbital parameters.

As an important part of our analysis, we first investigate the impact of the spin on eccentric, equatorial orbits, for both prograde and retrograde configurations. Secondly, we focus on the polar degree of freedom in the case of circular, off-equatorial geodesics, highlighting the emergence of the new type of resonances. Lastly, we estimate the role of the spin and of the polar degree of freedom on the number of cycles of the emitted GW.

The following simulations have been performed using the $\mathsf{Kerr Geodesics}$ package~\cite{BHPToolkit:KerrGeodesics} of the $\mathsf{BlackHolePerturbationToolkit}$~\cite{BHPToolkit}.


\subsection{Equatorial case\label{Sec: Numerical equatorial case}}

For equatorial orbits, we expect the system to undergo resonant peaks in the inner binary eccentricity of classes (i) and (ii), according to the classification of Sec.~\ref{Sec: Resonances classification}. These are the resonances that persist in the Schwarzschild limit, which provide an insightful comparison to the prograde and retrograde orbits in Kerr spacetime. 

\begin{figure}[ht]
    \centering
    \includegraphics[width=0.49\textwidth]{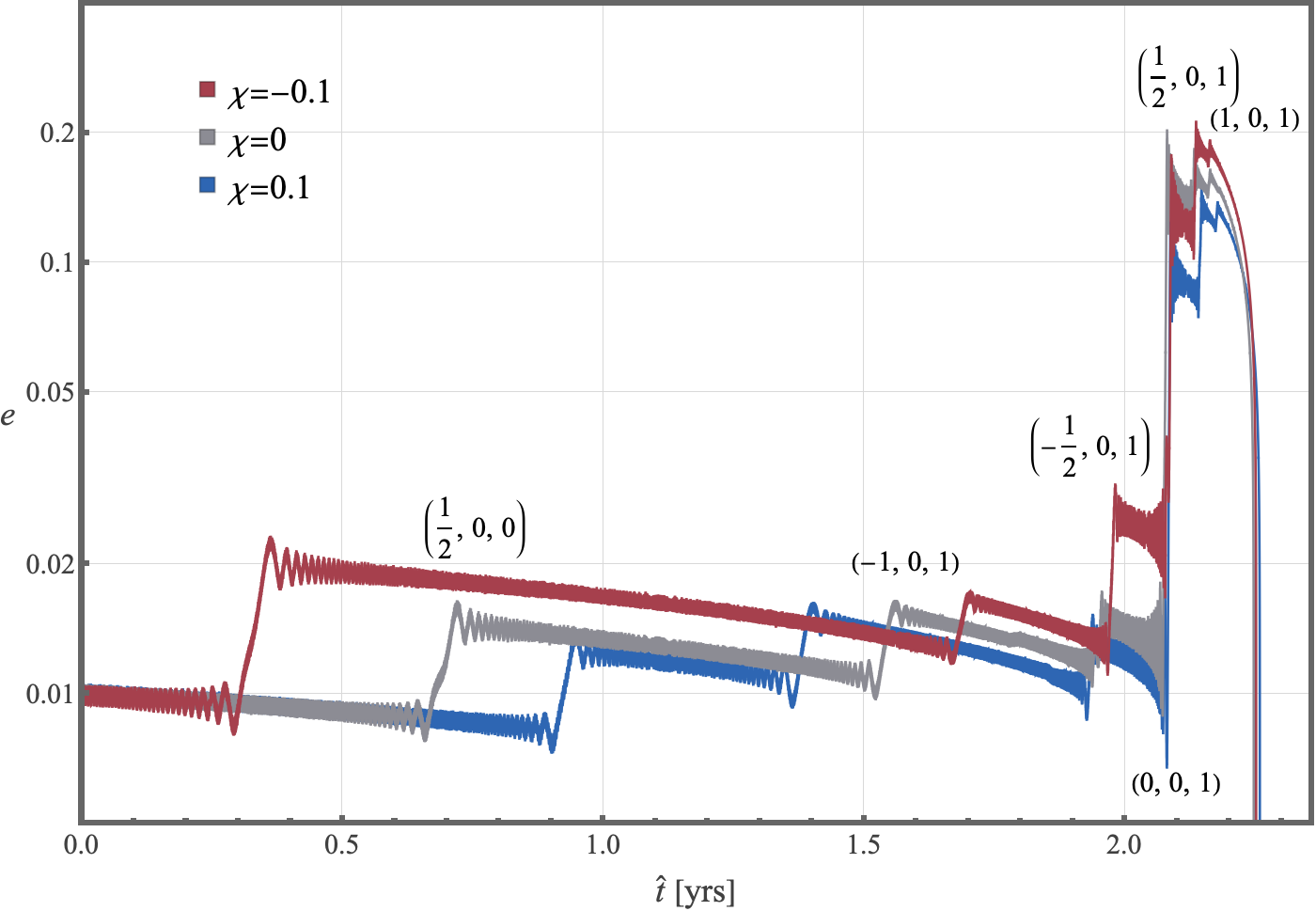}
    \begin{picture}(0,0)
        \tiny \put(-87,137){$(k,l,m)$}
    \end{picture}
    \caption{Inner binary's eccentricity $e$ as a function of time $\hat{t}$, comparing a retrograde orbit with $\chi=-0.1$ (red curve), a prograde orbit with $\chi=0.1$ (blue curve) and the Schwarzschild case for $\chi=0$ (grey curve).
    The parameters and initial conditions are: $M_*=8\times 10^6\, \rm M_{\odot}$, $m_1=20\,\rm M_{\odot}$, $m_2=10\, \rm M_{\odot}$, $\hat e=0.1$, $\hat z_1=0$, $\hat a=7.5 GM_*/c^2$, $a_0\sim 0.000427\, \text{AU}$, $e_0=0.01$, $\gamma_0=0^{\circ}$, $I_0=20^{\circ}$, $\vartheta_0=0^{\circ}$, $\hat\Psi_0=0^{\circ}$.
    For the chosen set of orbital parameters, the timescale assumptions are satisfied as  $T_{\text{inn,orb}}/T_{\text{out,orb}}\sim2\times 10^{-3}$, $T_{\text{inn,orb}}/T_{\text{tidal}}\sim4\times 10^{-3}$ and $T_{\text{inn,RR}}/T_{\text{out,RR}}\sim6\times 10^{-2}$.
    Numbers in parentheses indicate the values of $(k,l,m)$  corresponding to each resonance peak.
    It is important to note that only the resonances such that $\dot{\gamma}>0$ are matched. The GW emitted between the first and last resonance in the plot has peak frequency lying in the range $[0.04, 0.14]$ Hz.}
\label{Fig: equatorial_plot}
\end{figure}

In Fig.~\ref{Fig: equatorial_plot} we show the evolution of the eccentricity of the inner binary as a function of the asymptotic time for $\chi=-0.1$ (red curve), $\chi=0.1$ (blue curve) and $\chi=0$ (grey curve).
As expected, a nonzero spin shifts the timing of the resonances relative to the non-spinning case, reflecting the spin dependence of the fundamental frequencies of the outer orbit.  
For each resonance, prograde and retrograde orbits behave oppositely.
For example, the resonance $(1/2,0,0)$ occurs earlier for retrograde orbits and later for prograde ones, whereas the opposite happens for the resonance $(-1,0,1)$.
The zero-spin grey curve, consistently, always lies between the retrograde and prograde curves.

The different impact of the spin on prograde and retrograde orbits is reflected not only in the times at which resonances occur, but also in the amplitudes of the corresponding eccentricity peaks. However, comparing the amplitude of a given resonance for different values of the spin is not straightforward, since the orbital parameters of the inner binary generally differ at the time of resonance. A qualitative estimate of the spin dependence can be obtained by evaluating the corresponding Fourier coefficient as a function of the spin, keeping all other parameters fixed.


\subsection{Off-equatorial case\label{Sec: Numerical off-equatorial case}}

\subsubsection*{Polar resonances}

\begin{figure}[ht]
    \centering
\includegraphics[width=0.485\textwidth]{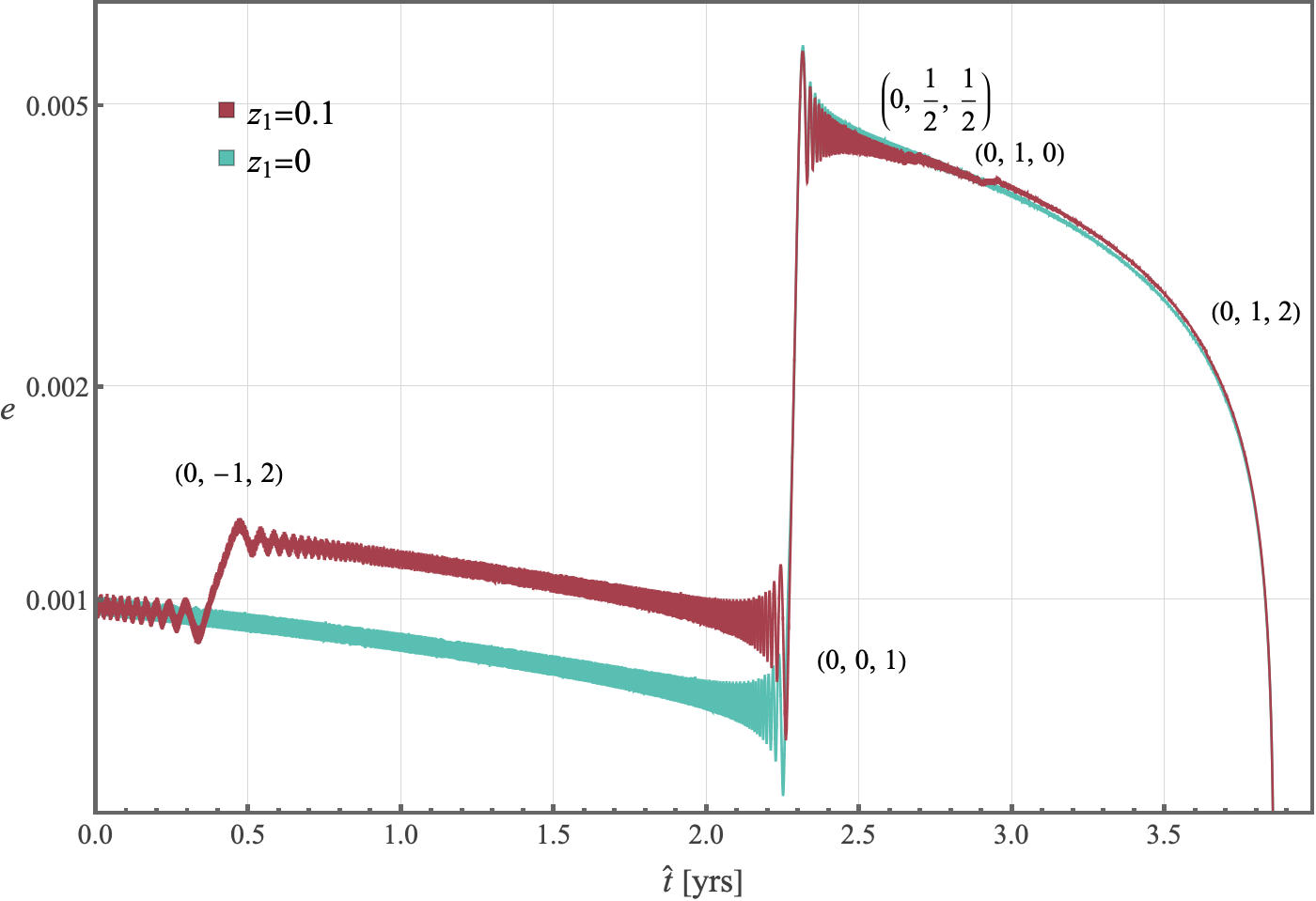}
    \begin{picture}(0,0)
        \tiny \put(-83,140){$(k,l,m)$}
    \end{picture}
    \caption{Inner binary's eccentricity $e$ as a function of time $\hat{t}$, comparing an inclined orbit with $\hat{z}_1=0.1$ (red curve) and equatorial configuration with $\hat{z}_1=0$ (cyan curve). 
    The parameters and initial conditions are: $M_*=3\times 10^7\, \rm M_{\odot}$, $m_1=20\,\rm M_{\odot}$, $m_2=10\, \rm M_{\odot}$, $\chi=-0.4$, $\hat e=0$, $\hat a=7.5 GM_*/c^2$, $a_0\sim 0.000486\, \text{AU}$, $e_0=0.001$, $\gamma_0=0^{\circ}$, $I_0=60^{\circ}$, $\vartheta_0=0^{\circ}$, $\hat\Psi_0=0^{\circ}$.
    For the chosen set of orbital parameters with $\hat{z}_1=0.1$ the timescale assumptions are satisfied as $T_{\text{inn,orb}}/T_{\text{out,orb}}\sim7\times 10^{-4}$, $T_{\text{inn,orb}}/T_{\text{tidal}}\sim7\times 10^{-4}$ and $T_{\text{inn,RR}}/T_{\text{out,RR}}\sim 10^{-2}$.
    Numbers in parentheses indicate the values of $(k,l,m)$ corresponding to each resonance peak.
    It is important to note that only the resonances such that $\dot{\gamma}(0)<k\Omega_{\hat{r}}+l\Omega_{\hat{z}}+m\Omega_{\hat{\Psi}}$, with $\dot{\gamma}>0$ are matched.
    For this reason, the polar resonance $(0,1/2,-1/2)$, whose condition is met at earlier times, does not appear in the plot.
    The peak GW frequency between the first and last resonance in the plot lies in the range $[0.03,0.11]$ Hz.}
\label{Fig: Polar_plot}
\end{figure}

To study how the eccentricity evolution of the inner binary is affected by the polar degree of freedom,
we focus on the class (iii), \textit{i.e.}, polar resonances. As confirmed by the numerical integration performed in Fig.~\ref{Fig: Polar_plot}, we observe polar resonant excitations only for off-equatorial geodesics (red curve), while they are absent in the equatorial case (cyan curve). 
Notice also that the numerical integration does not resolve the peak $(0,1,2)$ in Fig.~\ref{Fig: Polar_plot}. This feature is primarily related to the fact that the combination $\Omega_{\hat{z}}+2\Omega_{\hat{\Psi}}$ is met at the last stages of the inspiral. Here the semi-major axis of the inner binary reaches its lowest values and, since the coefficient $(0,1,2)$ scales as $a^{3/2}$, it is accordingly suppressed. Additionally, at this point of the inspiral, the precession frequency grows fast enough to damp the resonant transient. This feature also persists for different orbital configurations. By contrast, the other resonances occurring during the evolution give rise to appreciable peaks. In particular, the polar peak $(0,-1,2)$ can alter the evolution of the system with respect to the equatorial case, even for small values of $\hat{z}_1$, for a long portion of the inspiral. 


\subsubsection*{Radial-polar resonances}

\begin{figure}[ht]
    \centering
    \includegraphics[width=0.48\textwidth]{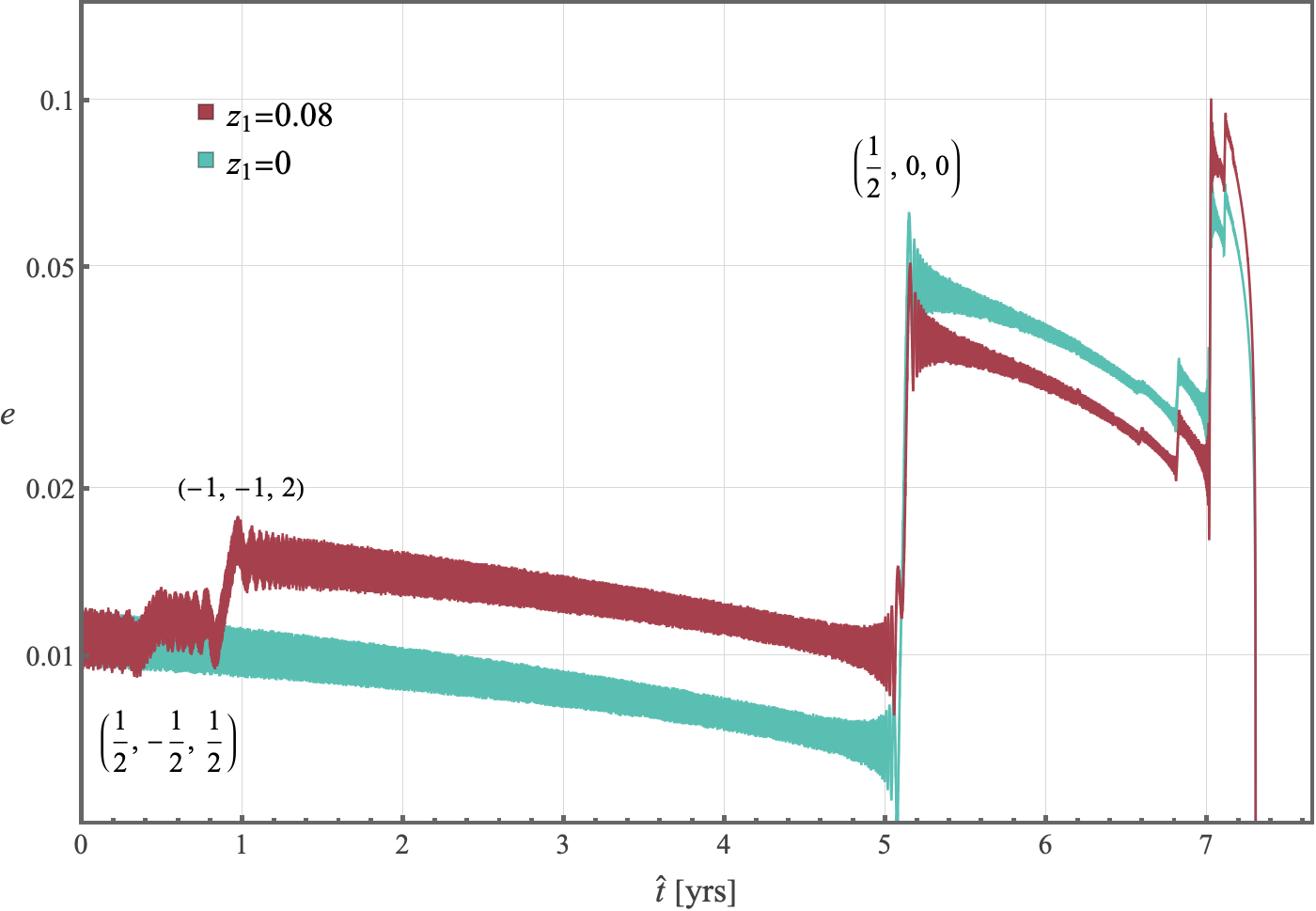}
    \begin{picture}(0,0)
        \tiny \put(-83,140){$(k,l,m)$}
    \end{picture}
    \caption{Inner binary's eccentricity $e$ as a function of time $\hat{t}$, comparing inclined eccentric orbit with $\hat{z}_1=0.08$ (red curve) and equatorial  eccentric orbit with $\hat{z}_1=0$ (cyan curve).
    The parameters and initial conditions are: $M_*=8\times 10^6\, \rm M_{\odot}$, $m_1=16\rm M_{\odot}$, $m_2=14\, \rm M_{\odot}$, $\chi=-0.6$, $\hat e=0.06$, $\hat a=9.7 GM_*/c^2$, $a_0\sim 0.0006\, \text{AU}$, $e_0=0.01$, $\gamma_0=0^{\circ}$, $I_0=60^{\circ}$, $\vartheta_0=0^{\circ}$, $\hat\Psi_0=0^{\circ}$. 
    For the chosen set of orbital parameters with $\hat{z}_1=0.08$ the timescales assumptions are satisfied as $T_{\text{inn,orb}}/T_{\text{out,orb}}\sim2\times 10^{-3}$, $T_{\text{inn,orb}}/T_{\text{tidal}}\sim5\times 10^{-3}$ and $T_{\text{inn,RR}}/T_{\text{out,RR}}\sim 10^{-1}$.
    Numbers in parentheses indicate the values of $(k,l,m)$ corresponding to each resonance peak. The emitted GW during the resonant phase of the plot has peak frequency that lies in the range $[0.02,0.09]$ Hz.}
\label{Fig: radialpolar_plot}
\end{figure}

To complete the numerical analysis, we investigate the more general case which includes both the radial and polar degrees of freedom, thereby inducing in principle all the resonance classes (i)-(iv). A representative example is depicted in Fig.~\ref{Fig: radialpolar_plot}.
We again consider positive values of $\dot{\gamma}$ and initiate the integration at a configuration that satisfies the underlying assumptions of the model. To highlight the role of radial-polar resonances, we stress the fact that, depending on the specific orbital configuration, these kinds of resonances can be the first to be encountered by the system during the inspiral. This feature is realized in Fig.~\ref{Fig: radialpolar_plot}, for the $(1/2,-1/2,1/2)$ and $(-1,-1,2)$ peaks (red curve). These resonances alter the dynamics of the inner binary for $\sim$4 yrs, until the binary encounters the next resonance. Even though their amplitudes are proportional to both the small parameters $\hat{e}$ and $\hat{z}_1$, neglecting the effect of radial-polar resonances would compromise the correct reconstruction of the GW signal emitted by the inner binary, as we will explore in the following subsection.


\subsection{Potential observability\label{Sec: Potential observability discussion}}

Precession resonances can potentially have an imprint in GW signals originating from b-EMRI systems. In fact, the resonant modulations of the eccentricity of the inner binary induce a modification of the frequency of the emitted GW signal. Even if this effect is small, it is accumulated over the large number of orbital cycles completed by the inner binary before merger. In the following, we provide a back-of-the-envelope computation of the number of GW cycles expected from the configurations examined in Figs.~\ref{Fig: equatorial_plot}, \ref{Fig: Polar_plot} and~\ref{Fig: radialpolar_plot}.

For elliptic motion, we use as a reference the peak frequency of the emitted GW, whose reliable estimate is given by~\cite{Hamers:2021eir}
\begin{equation} 
\label{eq:frequencypeak}
    f_{\mathrm{GW}}^{\text{peak}}= f_{\text{orb}}~n_{\text {peak}}(e).
\end{equation}
Here 
\begin{equation} 
\label{eq:f_orb}
    f_{\text{orb}}=\frac{1}{u^{\hat{t}}}\frac{1}{2 \pi} \sqrt{\frac{G M}{a^3}}
\end{equation}
denotes the orbital frequency of a circular binary of total mass $M$, and $n_{\text {peak}}(e)$ reads
\begin{equation}
    n_{\text {peak}}(e) \simeq 2\left(1+\sum_{k=1}^4 c_k e^k\right)\left(1-e^2\right)^{-3 / 2}
\end{equation}
where the fit coefficients $c_k$ are $c_1=-1.01678$, $c_2=5.57372$, $c_3=-4.9271$, and $c_4=1.68506$.

As a first remark, we stress that in every configuration analyzed here, the peak frequency lies inside the expected LISA frequency band $(0.001-0.1)~\mathrm{Hz}$ throughout most of the evolution.

To give a qualitative estimate of the dephasing caused by the precession resonances, we focus on the fundamental mode (2,2), rather than on the peak frequency.
For eccentric binaries, the (2,2) mode of the GW signal can be decomposed into harmonics with frequencies~\cite{Patterson:2024vbo}
\begin{equation}
    f_{\mathfrak{c}}=2f_\phi+ \mathfrak{c} f_r~, \quad \mathfrak{c}\in \mathbb{Z},
\end{equation}
where $f_\phi$ and $f_r$ are the azimuthal and radial frequencies, respectively, defined as 
\begin{equation}
   f_\phi= f_{\text{orb}}+\frac{1}{2\pi}\left\langle\frac{d\gamma}{d\hat{t}}\right\rangle,  \quad f_r= f_{\text{orb}}.
\end{equation}
Here, we consider the component $\mathfrak{c}=0$, which dominates in the quasi-circular regime, and we define the accumulated number of cycles during the inspiral as~\cite{Kavanagh:2020cfn, Li:2026uva,Maselli:2020zgv} 
\begin{equation}
N_\text{cycles}=\int_{\hat{t}_i}^{\hat{t}_f}f_{0}(\hat{t})\,d\hat{t}.
\end{equation}
We perform the integration over the interval from $\hat{t}_i=0$ to $\hat{t}_f=0.85 ~T_{\text{inn,RR}}$. This conservative cutoff allows us to compare the accumulated cycles, omitting the late-inspiral regime for which our model is less reliable.

In the case of the configuration examined in Fig.~\ref{Fig: equatorial_plot}, we compute the difference in the number of cycles between the zero-spin case and both the prograde and retrograde orbits. We have obtained $\Delta N_\text{cycles}\sim 7 \times 10^3$ and $\Delta N_\text{cycles}\sim -9 \times 10^3$ for the prograde and retrograde cases, respectively. The conservative detectability threshold commonly used is that a phase difference of order one radian would be observable by LISA~\cite{Li:2026uva,Maselli:2020zgv}. According to this criterion, our estimate suggests that even a small value of the BH spin can potentially leave detectable imprints on the GW signal. 

Similarly, the inclusion of the polar degree of freedom can lead to an appreciable dephasing. To confirm this statement, we proceed as before by computing the differences in the number of cycles between the inclined and the equatorial case for both the configurations analyzed in Figs.~\ref{Fig: Polar_plot} and~\ref{Fig: radialpolar_plot}. For the case of Fig.~\ref{Fig: Polar_plot}, we find $\Delta N_\text{cycles}\sim -2 \times 10^2$. This smaller value compared to the previous case is consistent with the fact that the system evolves through smaller values of the eccentricity $e$. Nonetheless, this result falls within the conservative criterion introduced above.
For the case of Fig.~\ref{Fig: radialpolar_plot}, we obtain $\Delta N_\text{cycles}\sim 20$, suggesting that also radial-polar resonances could be detectable by LISA. These results highlight the importance of further investigating resonances in b-EMRIs with the aim of improving the modeling of accurate GW templates.


\section{\label{Sec:Conclusions}Conclusions}

In this work, we studied precession resonances in compact binaries orbiting a Kerr SMBH in the
strong-gravity regime. We considered both equatorial and off-equatorial outer orbits. We found that, compared to the non-spinning Schwarzschild case previously examined in
Refs.~\cite{Cocco:2025adu,Cocco:2025udb},
the spin of the SMBH modifies both the fundamental frequencies of the outer motion and the tidal couplings, affecting when resonances occur and the amplitudes of the corresponding eccentricity excitations. For equatorial configurations, our numerical results show that the shifts relative to the Schwarzschild case depend on both the direction of the outer orbital motion (prograde or retrograde) and the specific resonance considered.

We have also derived compact analytical expressions for the resonance coefficients through an expansion in the outer orbital eccentricity and inclination, including mixed terms, while retaining the full dependence on the spin of the SMBH.

Allowing the outer orbit to depart from the equatorial plane reveals two additional types of resonances at the perturbative order considered in this paper: polar resonances, which arise even for circular outer orbits, and mixed radial–polar resonances, which require both nonzero outer eccentricity and inclination. Both types occur in the electric and magnetic tidal sectors. Their Fourier coefficients vanish in the zero-spin limit, showing that these additional resonances are specific to off-equatorial motion around a rotating black hole.

Furthermore, our numerical examples demonstrate that even a small outer inclination can modify the eccentricity evolution of the inner binary over an extended portion of the inspiral phase. In particular, mixed radial–polar resonances can be among the first encountered by the system and therefore influence the subsequent evolution. These results show that the additional resonance families that we found can have an impact beyond the immediate resonant encounter.

Our estimates of the accumulated GW cycles suggest that both spin-dependent and off-equatorial tidal effects may be relevant for waveform modeling. These findings motivate the development of gravitational waveforms incorporating precession resonances and a dedicated assessment of whether their signatures can be distinguished by LISA.

An immediate extension of this work is to push the perturbative expansion in $\hat{e}$ and $\hat{z}_1$ to higher orders. This will most likely give rise to additional resonances, as suggested by the numerical analysis of Sec.~\ref{Sec:Numerical} (see also Refs.~\cite{Cocco:2025adu,Cocco:2025udb}).

Moreover, we found most of the assumptions listed in Sec.~\ref{Sec:Assumptions} to be satisfied over a wide range of orbital parameters, the most restrictive one being that GW emission from the outer orbit can be neglected. Relaxing this assumption would not modify the resonance conditions, which depend on the instantaneous fundamental frequencies, but would affect the time evolution of the system, as the outer frequencies would slowly drift during the inspiral.
Radiation reaction for the outer orbit could be included, for instance, through the method of osculating geodesics~\cite{Pound:2007th}, in which the worldline is described at each instant by the geodesic tangent to it, with orbital elements evolving under the effect of a perturbing force. Since no averaging over the outer orbit is involved, the outer phases entering the resonance conditions remain fully resolved. This would allow the model to probe a larger region of the parameter space. 

Conversely, GW emission from the inner binary was modeled through the leading-order (2.5PN) dissipative contribution~\eqref{Petersequations}. While accurate in the early inspiral, this description becomes less reliable as the binary approaches merger. Including higher-order PN dissipative corrections would improve the modeling of the late inspiral, and hence the predicted timing of the resonances and the resulting dephasing of the GW signal.

Finally, an important direction is the construction of GW templates that include the effect of precession resonances, together with a more refined estimate of the dephasing of the emitted signal.


\section*{Acknowledgments}

M.C., T.H., M.O. and D.P. acknowledge support by the “Center of Gravity”, which is
a Center of Excellence funded by the Danish National Research Foundation under grant No. 184. 
M.C., G.G., F.M., M.O. and D.P. acknowledge financial support from
the Italian Ministry of University and Research (MUR) through the program “Dipartimenti di Eccellenza 2018-2022” (Grant SUPER-C). G.G. and M.O. acknowledge support from “Fondo Ricerca d’Ateneo” of the University of Perugia. 
M.C. acknowledges financial support from the European Research Council (ERC) grant (EFTGrav, Grant Agreement No. 101220945).

\newpage
\clearpage
\newpage

\appendix

\section{\label{Appendix A: Tidal moments}Tidal moments}

We provide the explicit components of the electric $\mathscr{E}_{ij}$ and magnetic $\mathscr{B}_{ij}$ tidal tensors. These results are valid for arbitrary time-like geodesics in the Kerr black hole spacetime, as derived in Ref.~\cite{Camilloni:2023rra}.

It is convenient to define the following auxiliary quantities
\begin{equation}
\label{Auxiliarydefinitionstidalmoments1}
    S = \hat{r}^2 + \hat{K} \ , \quad T = \hat{K} - \mathrm{a}^2 \hat{z}^2\ .
\end{equation}
and
\begin{equation}
\label{Auxiliarydefinitionstidalmoments2}
    I_1 = \frac{G M_*}{c^2}\frac{ \hat{r}}{\Sigma^3} \left( \hat{r}^2 \!- 3 \mathrm{a}^2 \hat{z}^2 \right),\!\! \quad
    I_2 = \frac{G M_*}{c^2}\frac{\mathrm{a} \hat{z}}{\Sigma^3} \left( 3 \hat{r}^2 \!- \mathrm{a}^2 \hat{z}^2 \right)  .
\end{equation}
Notice the useful identity $\Sigma=S-T$.
\begin{widetext}
\noindent
Electric tidal moments read
\begin{equation}
\label{Explicitelectrictidal}
\begin{aligned}
\mathscr{E}_{11} &= \left[ 1 - \frac{3ST}{\hat{K}\Sigma^2} (\hat{r}^2 - \mathrm{a}^2 \hat{z}^2) \cos^2 \hat{\Psi} \right] I_1 + \frac{6ST}{\hat{K}\Sigma^2} \mathrm{a} \hat{r} \hat{z} \cos^2 \hat{\Psi} I_2 \ , \\
\mathscr{E}_{12} &=\mathscr{E}_{21} =-\frac{3ST}{\hat{K}\Sigma^2} \left[ (\hat{r}^2 - \mathrm{a}^2 \hat{z}^2) I_1 - 2 \mathrm{a} \hat{r} \hat{z} I_2 \right] \sin \hat{\Psi} \cos \hat{\Psi} \ , \\
\mathscr{E}_{13} &= \mathscr{E}_{31} =-\frac{3\sqrt{ST}}{\hat{K}\Sigma^2} \left[ \mathrm{a} \hat{r} \hat{z} (S + T) I_1 + (\hat{r}^2 T - \mathrm{a}^2 S \hat{z}^2) I_2 \right] \cos \hat{\Psi} \ , \\
\mathscr{E}_{22} &= \left[ 1 - \frac{3ST}{\hat{K}\Sigma^2} (\hat{r}^2 - \mathrm{a}^2 \hat{z}^2) \sin^2 \hat{\Psi} \right] I_1 + \frac{6ST}{\hat{K}\Sigma^2} \mathrm{a} \hat{r} \hat{z} \sin^2 \hat{\Psi} I_2 \ , \\
\mathscr{E}_{23} &=\mathscr{E}_{32} = -\frac{3\sqrt{ST}}{\hat{K}\Sigma^2} \left[ \mathrm{a} \hat{r} \hat{z} (S + T) I_1 + (\hat{r}^2 T - \mathrm{a}^2 S \hat{z}^2) I_2 \right] \sin \hat{\Psi} \ , \\
\mathscr{E}_{33} &= \left[ 1 + \frac{3}{\hat{K}\Sigma^2} (\hat{r}^2 T^2 - \mathrm{a}^2 S^2 \hat{z}^2) \right] I_1 - \frac{6ST}{\hat{K}\Sigma^2} \mathrm{a} \hat{r} \hat{z} I_2 \ .
\end{aligned}
\end{equation}
Magnetic tidal moments were explicitly derived, in the most general case of off-equatorial orbits in Kerr spacetime, in Ref.~\cite{Camilloni:2023rra}, and can be written as
\begin{equation}
\label{Explicitmagnetictidal}
\begin{aligned}
\mathscr{B}_{11} &= \frac{6ST}{\hat{K}\Sigma^2} \mathrm{a}\hat{r}\hat{z} \cos^2 \hat{\Psi} I_1 - \frac{1}{\hat{K}\Sigma^2} \left[ (\hat{r}^2 T + \mathrm{a}^2 S \hat{z}^2)(S - T) - 3ST \left( \hat{r}^2 - \mathrm{a}^2 \hat{z}^2 \right) \cos^2 \hat{\Psi} \right] I_2 \ ,\\
\mathscr{B}_{12} &=\mathscr{B}_{21} = \frac{3ST}{\hat{K}\Sigma^2} \left[ 2\mathrm{a}\hat{r}\hat{z}I_1 + \left( \hat{r}^2 - \mathrm{a}^2 \hat{z}^2 \right) I_2 \right] \sin \hat{\Psi} \cos \hat{\Psi} \ , \\
\mathscr{B}_{13} &=\mathscr{B}_{31} = -\frac{3\sqrt{ST}}{\hat{K}\Sigma^2} \left[ (\hat{r}^2 T - \mathrm{a}^2 S \hat{z}^2) I_1 - \mathrm{a}\hat{r}\hat{z}(S + T)I_2 \right] \cos \hat{\Psi} \ , \\
\mathscr{B}_{22} &= \frac{6ST}{\hat{K}\Sigma^2} \mathrm{a}\hat{r}\hat{z} \sin^2 \hat{\Psi} I_1 + \frac{1}{\hat{K}\Sigma^2} \left[ \hat{r}^2 T (2S + T) - \mathrm{a}^2 \hat{z}^2 S (S + 2T) - 3ST \left( \hat{r}^2 - \mathrm{a}^2 \hat{z}^2 \right) \cos^2 \hat{\Psi} \right] I_2 \ ,\\
\mathscr{B}_{23} &=\mathscr{B}_{32} = -\frac{3\sqrt{ST}}{\hat{K}\Sigma^2} \left[ (\hat{r}^2 T - \mathrm{a}^2 S \hat{z}^2) I_1 - \mathrm{a}\hat{r}\hat{z}(S + T)I_2 \right] \sin \hat{\Psi} \ , \\
\mathscr{B}_{33} &=-\frac{6  S T}{\hat{K} \Sigma^2}  \mathrm{a}\hat{r} \hat{z} I_1+ \frac{1}{\hat{K} \Sigma^2} \left[ -\hat{r}^2 T (S + 2T) + \mathrm{a}^2 S (2S + T) \hat{z}^2 \right]I_2\ .
\end{aligned}
\end{equation}
As described in Sec.~\ref{sec:Tidal Hamiltonian}, these components are used to construct the tidal Hamiltonian $H_{\text{tidal}}$, which encodes both electric-type and magnetic-type interactions. We provide here the orbit-averaged, implementation-ready expressions for the two Hamiltonians
\begin{equation}
\label{Fullelectrichamiltonian}
\begin{aligned}
\langle H_{\mathscr{E}}\rangle&=-\mu c^2\frac{a^2}{16 \hat{K} \Sigma^2} \left[ 
4 \Big( I_1 \hat{K} \Sigma^2 + 3 S T (-I_1 \hat{r}^2 + 2 \mathrm{a} I_2 \hat{r} \hat{z} + \mathrm{a}^2 I_1 \hat{z}^2) \cos^2 \hat{\Psi} \Big) \right. \\
& \quad \times \Big( (e^2 - 1) (\cos \vartheta \sin \gamma + \cos \gamma \cos I \sin \vartheta)^2 - (1 + 4e^2) (\cos \gamma \cos \vartheta - \cos I \sin \gamma \sin \vartheta)^2 \Big) \\
& + 2 \Big( -6 \mathrm{a} I_2 \hat{r} S T \hat{z} + I_1 (3 \hat{r}^2 T^2 - 3 \mathrm{a}^2 S^2 \hat{z}^2 + \hat{K} \Sigma^2) \Big) (-2 - 3e^2 + 5e^2 \cos 2\gamma) \sin^2 I \\
& + 6 \sqrt{ST} \Big( I_2 \hat{r}^2 T + \mathrm{a} I_1 \hat{r} (S + T) \hat{z} - \mathrm{a}^2 I_2 S \hat{z}^2 \Big) \cos \hat{\Psi} \\
& \quad \times \Big( 10e^2 \cos \vartheta \sin 2\gamma \sin I + (-2 - 3e^2 + 5e^2 \cos 2\gamma) \sin \vartheta \sin 2I \Big) \\
& + 3 S T \Big( -2 \mathrm{a} I_2 \hat{r} \hat{z} + I_1 (\hat{r} - \mathrm{a}\hat{z})(\hat{r} + \mathrm{a}\hat{z}) \Big) \cos \hat{\Psi} \sin \hat{\Psi} \\
& \quad \times \Big( 20e^2 \cos 2\vartheta \cos I \sin 2\gamma + \sin 2\vartheta \big( 5e^2 \cos 2\gamma (3 + \cos 2I) + 2(2 + 3e^2) \sin^2 I \big) \Big) \\
& + 6 \sqrt{ST} \Big( I_2 \hat{r}^2 T + \mathrm{a} I_1 \hat{r} (S + T) \hat{z} - \mathrm{a}^2 I_2 S \hat{z}^2 \Big) \sin \hat{\Psi} \\
& \quad \times \Big( 10e^2 \sin 2\gamma \sin \vartheta \sin I + (2 + 3e^2 - 5e^2 \cos 2\gamma) \cos \vartheta \sin 2I \Big) \\
& - 4 \Big( (1 + 4e^2) (\cos \vartheta \cos I \sin \gamma + \cos \gamma \sin \vartheta)^2 - (e^2 - 1) (\cos \gamma \cos \vartheta \cos I - \sin \gamma \sin \vartheta)^2 \Big) \\
& \quad \times \left. \Big( I_1 \hat{K} \Sigma^2 + 3 S T (-I_1 \hat{r}^2 + 2 \mathrm{a} I_2 \hat{r} \hat{z} + \mathrm{a}^2 I_1 \hat{z}^2) \sin^2 \hat{\Psi} \Big) \right]
\end{aligned}
\end{equation}
\begin{equation}
\label{Fullmagnetichamiltonian}
\begin{aligned}
\langle H_{\mathscr{B}}\rangle&=\mu c\frac{3 (m_1 - m_2) a e \sqrt{G M a (1-e^2)} }{4 M \hat{K} \Sigma^2}  \Big[ \Big. \sin \gamma \Big\{ \Big( I_2 \hat{r}^2 T (S + 2T) + 6 \mathrm{a} I_1 \hat{r} S T \hat{z} - \mathrm{a}^2 I_2 S (2S + T) \hat{z}^2 \\
& \quad + S T (-I_2 \hat{r}^2 - 2 \mathrm{a} I_1 \hat{r} \hat{z} + \mathrm{a}^2 I_2 \hat{z}^2) \cos(2(\vartheta - \hat{\Psi})) \Big) \sin 2I + 4 \sqrt{ST} (-I_1 \hat{r}^2 T + \mathrm{a} I_2 \hat{r} (S + T) \hat{z} \\
&+ \mathrm{a}^2 I_1 S \hat{z}^2) \cos 2I \sin(\vartheta - \hat{\Psi}) \Big\}   + 2 \cos \gamma \Big\{ -2 \sqrt{ST} (-I_1 \hat{r}^2 T + \mathrm{a} I_2 \hat{r} (S + T) \hat{z} + \mathrm{a}^2 I_1 S \hat{z}^2) \cos I \cos(\vartheta - \hat{\Psi})\\
&+ S T (-I_2 \hat{r}^2 - 2 \mathrm{a} I_1 \hat{r} \hat{z} + \mathrm{a}^2 I_2 \hat{z}^2) \sin I \sin(2(\vartheta - \hat{\Psi})) \Big\} \Big. \Big]
\end{aligned}
\end{equation}
\end{widetext}


\section{\label{Appendix C: Results of the Double Expansion}Perturbative double expansion}

Here we present the expanded forms of the quantities introduced in Sec.~\ref{Sec: Perturbative approach} up to perturbative order $\mathscr{O}(\hat{e},\hat{z}_1)$.\\

For the oscillating functions we get
\begin{equation} \label{Expansionoscillatingfunctions} \begin{aligned} \varphi_{\hat{t}\hat{r}}\! =& \hat{e} \frac{2 \hat{\sigma} (\chi^3 \!+\! \chi^2 (\hat{\sigma}\!-\!1) \sqrt{\hat{\sigma}} \!+\! \chi \hat{\sigma}^2 \!+\! ( \hat{\sigma}\!-\!3) \hat{\sigma}^{5/2}) }{\!\!\sqrt{\!-3 \chi^2 \!+\! 8 \chi\! \sqrt{\hat{\sigma}}\! +\! (\hat{\sigma}\!-\!6) \hat{\sigma}} ((2\!-\!\hat{\sigma}) \hat{\sigma}\!-\!\chi^2)}\\&\hspace{0cm}\times\frac{G M_*}{c^3}\!\sin q_{\hat{r}},\\ \varphi_{\hat{t}\hat{z}} \!=& 0\,,\\ \varphi_{\hat{r}\hat{r}} \!=&\hat{\sigma} \frac{G M_*}{c^2}- \hat{e}\,\hat{\sigma} \frac{G M_*}{c^2} \cos q_{\hat{r}}\,,\\ \varphi_{\hat{z}\hat{z}} \!=& \hat{z}_1 \sin q_{\hat{z}}\,,\\ \varphi_{\hat{\Psi}\hat{r}} \!=& \hat{e}\, \frac{2 ( \sqrt{\hat{\sigma}}\!-\!\chi)^2 \hat{\sigma}^{1/4} }{\chi^2 \!+\! (\hat{\sigma}\!-\!2) \hat{\sigma}}\sqrt{\frac{2 \chi + ( \hat{\sigma}-3) \sqrt{\hat{\sigma}}}{\!-3 \chi^2 \!+\! 8 \chi \sqrt{\hat{\sigma}} \!+\! ( \hat{\sigma}\!-\!6) \hat{\sigma}}} \sin q_{\hat{r}}\,,\\ \varphi_{\hat{\Psi}\hat{z}}\!=&0\,. \end{aligned} 
\end{equation}
It is important to note that the functions $\varphi_{A B}$ are defined with respect to Mino time and, therefore, do not reduce to the functions $f_{A B}$ presented in Refs.~\cite{Cocco:2025udb, Cocco:2025adu} in the Schwarzschild limit $\chi \to 0$.

We then present the fundamental frequencies within the action-angle formalism, first those defined with respect to Mino time, $\Upsilon_A$, then those defined with respect to coordinate time, $\Omega_A$.
The first set of frequencies reads
\begin{equation}
\label{Expansionfundamentalfrequenciesmino}
\begin{aligned}
    \Upsilon_{\hat{t}} &=\frac{\hat{\sigma}^{5 / 4}\left(\chi+\hat{\sigma}^{3 / 2}\right)}{\sqrt{2 \chi+(\hat{\sigma}-3) \sqrt{\hat{\sigma}}}}\left(\frac{GM_*}{c^2}\right)^2\!,\\
    \Upsilon_{\hat{r}} &= \hat{\sigma}^{1/4} \sqrt{\frac{-3 \chi^2 + 8 \chi \sqrt{\hat{\sigma}} + (\hat{\sigma} - 6) \hat{\sigma}}{2 \chi + (\hat{\sigma} - 3) \sqrt{\hat{\sigma}}}}\frac{GM_*}{c}\,,\\
    \Upsilon_{\hat{z}}&= \hat{\sigma}^{1/4} \sqrt{\frac{3 \chi^2 - 4 \chi \sqrt{\hat{\sigma}} + \hat{\sigma}^2}{2 \chi + (\hat{\sigma} - 3) \sqrt{\hat{\sigma}}}} \frac{GM_*}{c}\,,\\
    \Upsilon_{\hat{\Psi}} &= \sqrt{\hat{\sigma}}\frac{GM_*}{c}\,.
\end{aligned}
\end{equation}
The set of fundamental frequencies in the asymptotic time is
\begin{equation}
\label{Expansionfundamentalfrequenciesasy}
\begin{aligned}
    \Omega_{\hat{r}} &=
    \frac{\sqrt{-3 \chi^2+8 \chi \sqrt{\hat{\sigma}}-6 \hat{\sigma}+\hat{\sigma}^2}}{\chi \hat{\sigma}+\hat{\sigma}^{5 / 2}} \frac{c^3}{G M_*}\,,\\
    \Omega_{\hat{z}} &= 
    \frac{\sqrt{3 \chi^2 - 4 \chi \sqrt{\hat{\sigma}} + \hat{\sigma}^2}}{\chi \hat{\sigma} + \hat{\sigma}^{5/2}} \frac{c^3}{G M_*}\,,\\
    \Omega_{\hat{\Psi}} &= \frac{\sqrt{2 \chi+(\hat{\sigma}-3) \sqrt{\hat{\sigma}}}}{\hat{\sigma}^{3 / 4}\left(\chi+\hat{\sigma}^{3 / 2}\right)} \frac{c^3}{G M_*}\,.
\end{aligned}
\end{equation}
We have verified that these frequencies reduce to their Schwarzschild counterparts~\cite{Cocco:2025adu} in the $\chi\to0$ limit. We have also checked that these frequencies reduce to the Newtonian ones in the weak field limit, namely $\Omega_{\hat{r},\hat{z},\hat{\Psi}}\to\Omega_N=\sqrt{G M_*/\hat{a}^3}$ for $\hat{\sigma}\to\infty$.\\
For completeness, we also provide the expanded forms of all three constants of motion in Kerr spacetime.
\begin{equation}
\label{Expandedconstantsofmotion}
\begin{aligned}
    \hat{E}&=\frac{\chi + (\hat{\sigma} - 2) \sqrt{\hat{\sigma}}}{\sqrt{2 \chi + (\hat{\sigma} - 3) \sqrt{\hat{\sigma}}} \hat{\sigma}^{3/4}} \,,\\
    \hat{L}_{\hat{z}}&=\frac{(\hat{\sigma}^2-2\sqrt{\hat\sigma}\chi+\chi^2)}{\sqrt{2 \chi + (\hat{\sigma} - 3) \sqrt{\hat{\sigma}}} \hat{\sigma}^{3/4}} \frac{G M_*}{c^2}\,,\\
    \hat{K}&=(\hat{L}_{\hat{z}}-\mathrm{a}\hat{E})^2=\frac{(\chi - \sqrt{\hat{\sigma}})^2 \hat{\sigma}^{3/2}}{2 \chi + (\hat{\sigma} - 3) \sqrt{\hat{\sigma}}} \left(\frac{G M_*}{c^2}\right)^2\,.
\end{aligned}
\end{equation}
It is worth noting that the frequencies (\ref{Expansionfundamentalfrequenciesmino}), (\ref{Expansionfundamentalfrequenciesasy}) and the constants of motion (\ref{Expandedconstantsofmotion}) do not depend on the perturbative parameters $\hat{e}$ and $\hat{z}_1$ at leading order in the expansion.

\setlength{\bibsep}{-0.6pt}
\bibliography{biblio}

\end{document}